\documentclass[aps,twocolumn,pra,groupedaddress,superscriptaddress,showkeys]{revtex4-2}

\usepackage[english]{babel}

\usepackage{float}
\usepackage[table,dvipsnames]{xcolor}
\usepackage{dcolumn}
\usepackage{bm}
\usepackage{graphicx}
\usepackage{amsmath}
\usepackage{amsfonts}
\usepackage{amssymb}
\usepackage{amsthm}
\usepackage{amstext}
\usepackage{amsbsy}
\usepackage{amsopn}
\usepackage{amscd}
\usepackage{amsxtra}
\usepackage{soul}

\usepackage{hyperref}
\usepackage{booktabs}
\usepackage{multirow}

\newcolumntype{C}{>{$}c<{$}}
\AtBeginDocument{
\heavyrulewidth=.08em
\lightrulewidth=.05em
\cmidrulewidth=.03em
\belowrulesep=.65ex
\belowbottomsep=0pt
\aboverulesep=.4ex
\abovetopsep=0pt
\cmidrulesep=\doublerulesep
\cmidrulekern=.5em
\defaultaddspace=.5em}

\newcommand{\Mc}[1]{\mathcal{#1}}
\newcommand{\setZ}{\mathbb{Z}}
\newcommand{\setN}{\mathbb{N}}
\newcommand{\setR}{\mathbb{R}}

\newcommand{\setC}{\mathbb{C}}

\newcommand{\bra}[1]{\langle #1 |}
\newcommand{\ket}[1]{|#1\rangle }
\newcommand{\braket}[2]{\langle #1| #2 \rangle }
\newcommand{\ii}{i}
\newcommand{\A}{\hat a}
\newcommand{\Ad}{\hat a^\dagger}
\newcommand{\B}{\hat b}
\newcommand{\Bd}{\hat b^\dagger}

\everymath{\displaystyle}

\begin{document}

\title{Enhancing quantum exchanges between two oscillators}

\author{Q. Ansel}
\email{quentin.ansel@univ-fcomte.fr}
\affiliation{Institut UTINAM, UMR 6213, CNRS, Universit\'{e} Bourgogne Franche-Comt\'{e}, Observatoire des Sciences de l'Univers THETA, 41 bis avenue de l'Observatoire, F-25010 Besan\c{c}on, France}

\author{A. D. Chepelianskii}
\affiliation{LPS, Université Paris-Saclay, CNRS, UMR 8502, F-91405, Orsay, France}


\author{J. Lages}
\affiliation{Institut UTINAM, UMR 6213, CNRS, Universit\'{e} Bourgogne Franche-Comt\'{e}, Observatoire des Sciences de l'Univers THETA, 41 bis avenue de l'Observatoire, F-25010 Besan\c{c}on, France}


\begin{abstract}
We explore the extent to which two quantum oscillators can exchange 
their quantum states efficiently through a three-level system which can be spin levels of 
colored centers in solids. High transition probabilities are obtained 
using Hamiltonian engineering and quantum control techniques. Starting from a weak 
coupling approximation, we derive conditions on the spin-oscillator interaction Hamiltonian that enable a high-fidelity exchange of quanta. We find that these conditions cannot be fulfilled for arbitrary 
spin-oscillator coupling. To overcome this limitation, we illustrate how a time-dependent control field applied to the three-level system can lead to an effective dynamic that performs the desired exchange of excitation. In the (ultra) strong coupling regime, an important loss of fidelity is induced by the dispersion of the excitation onto many Fock states of the oscillators. We show that this detrimental effect can be substantially reduced by suitable control fields, which are computed with optimal control numerical algorithms.
\end{abstract}

\maketitle

\section{Introduction}

The transfer of quantum information between different systems plays a central role in quantum technologies~\cite{kimble_quantum_2008,Ma_Quantum_2019,Fukui_2022,knill_scheme_2001}. Quantum oscillators (such as the modes of the electromagnetic field) are promising physical systems for the transport and storage of quantum information. One of the main issues is to manipulate/transfer this information efficiently since two quantum oscillators usually interact together through another quantum system. Engineering this interaction is therefore a key aspect to produce efficient quantum devices. This issue has been recently investigated with two coupled oscillators in~\cite{rosenblum_cnot_2018,gao_programmable_2018,zhang_engineering_2019}.

Engineering communication between oscillators to produce high-fidelity quantum operations requires accurate modeling of the physical system. Recent studies \cite{zhang_engineering_2019,Strauch_2010_Arbitrary,Yang_2013_Generating,Wang_2011_Deterministic} have focused on the interaction between photonic states of superconducting cavities mediated by transmons (or superconducting qubits). Similar interactions can be obtained by replacing the transmon with a few energy levels of one or several atoms. In this context, Nitrogen-Vacancy (NV) centers in diamond, or more generally solid-state spins, are interesting candidates for quantum technologies, due to their long coherence time~{\cite{Weber_Quantum_computing_2010,kurizki_quantum_2015,awschalom_quantum_2018}}. Solid-state spins can be coupled to several modes of the electromagnetic field (photons) or coupled to a vibrational mode (phonons)~\cite{Lee_2017,Teissier_strain_2014,Rosenfeld_2021_Efficient}.  

In this paper, we propose to revisit the topic with a model system of two harmonic oscillators, of different frequencies, coupled with a three-level system whose properties are similar to a spin-1 particle. This system can be experimentally realized, using a specific subset of three energy levels of solid-state spins~\cite{PhysRevB.91.121201,PhysRevA.92.020301,Lesik_Magnetic_measurements_2019,Bayliss_Optically_addressable_2020,stern_room-temperature_2022}. Oscillators are modes of either the electromagnetic field or a vibrational 
mode changing the fine structure tensor. Unlike the two-level systems, the three-level system has different frequency resonances that give rise to one or two excitation exchanges. We go beyond the Jaynes-Cummings model, and we consider a Rabi-type Hamiltonian~\cite{Xie_2017,Leggett_1987_Dynamics,Chilingaryan_2015}. This allows us to cover both the weak and the strong coupling regime~\cite{Xie_2017,PhysRevLett.105.140502,PhysRevLett.105.237001} (this latter is also qualified of "ultra-strong" or "deep-strong" when the physical system has other characteristic frequencies). The weak coupling regime is commonly realized in experiments~\cite{kurizki_quantum_2015,awschalom_quantum_2018}, but reaching the strong coupling can be interesting to speed up the interaction~\cite{kurizki_quantum_2015}. In order to improve the transfer protocol, we assume that the three-level system can be controlled by means of an additional classical field~\cite{MacQuarrie_Mechanical_2013}, such as a driven magnetic field whose coherence time is negligible. Note that a similar device is considered in Ref.~\cite{zhang_engineering_2019}, where the oscillators are coupled with a transmon, and a similar problem has been considered in \cite{PhysRevA.71.023805} with a Stimulated Raman adiabatic passage (STIRAP).

Different complementary techniques can be implemented to transfer $n$-quantum excitations between oscillators. The first one consists of manipulating the system parameters (cf. Hamiltonian engineering~\cite{Ma_Quantum_2019}) to design an eigenstate that produces exactly a Rabi-oscillation between $n$-excitations states. This simple strategy, which may have non-trivial solutions, only works well in the weak coupling regime. The second technique provides assistance when the first solution fails. The idea is to employ a control field that effectively produces the Rabi-oscillation. With the help of an averaged Hamiltonian~\cite{maricq_application_1982,brinkmann_introduction_2016} and with numerical optimal-control techniques, we determine control fields that maximize the efficiency of some transfer processes. Optimal Control Theory
(OCT)~\cite{brif_control_2010,bonnard_optimal_2012,glaser_training_2015,Boscain_Introduction_2021} has been an unavoidable tool to derive suitable field shapes for complex quantum control problems. It has been applied with success in a variety of studies in the field of quantum technologies. Closely related to our present work, we have the quantum control of bosonic modes with superconducting circuits~\cite{Ma_Quantum_2019}, the optimal engineering of beamsplitter~\cite{basilewitsch_engineering_2021}, the time-optimal control of driven spin-boson systems \cite{Jirari_Quantum_2006,jirari_quantum_2019,Jirari_time_2020}, and the optimal control of NV-centers~\cite{Rembold_Introduction_2020,Ansel_exploring_2022,Ansel_Optimal_2022}.

This paper is organized as follows. The model system is described in Sec.~\ref{sec:model_syst} and several of its key physical properties are presented. In Sec.~\ref{sec:weak_coupling}, we show how one and two photons exchanges can be realized efficiently in the weak coupling regime. In Sec.~\ref{sec:effective_hamiltonian}, we illustrate how time-dependent control fields applied to the three-level system can generate an effective Hamiltonian that performs the excitation exchange when the initial Hamiltonian cannot. In Sec.~\ref{sec:strong_coupling}, we consider the strong coupling regime and we explore the extent to which a  suitable control can limit the dispersion of the quantum excitation. Conclusion and prospective views are given in Sec.~\ref{sec:conclusion}. Some technical details are reported in the appendices.

\section{The Model System}
\label{sec:model_syst}

In this section, we present the physical model. First, we provide the main definitions, and then, we discuss the role of the environment. Finally, we present the physical effects that play a key role in our transfer protocols.

\subsection{General Definitions}

We consider a quantum system composed of two harmonic oscillators, noted respectively $a$ and $b$, in interaction with a three-level quantum system $S$ \footnote{I.e. here a spin-1 whose algebra of operator is identified with the $\mathfrak{su}(2)$ Lie-algebra in the representation $j=1$.}. We also consider a possible coherent control of the three-level system. The control can be assimilated with a modulated electromagnetic wave. 

The Hilbert space of the system is $\Mc H = \setC^3 \otimes \Mc F \otimes \Mc F$ with $\Mc F$ a bosonic Fock-space. We denote by $\ket{m_S,k_a,l_b}$ the states with $m\in\{1,0,-1\}$ the eigenvalues of $\hat S_z \in \mathfrak{su}(2)$, $k \in \setN$ the number of quanta in the oscillator $a$, and $l \in \setN$ the number of quanta in the oscillator $b$. In the notation of the state, the subscripts $S,a,b$ are used to recall the association of quantum numbers with the physical system. Dynamics of the system are given by the Schr\"odinger equation $d_t\ket{\psi(t)} = -\ii \hat H(t) \ket{\psi(t)}$, in $\hbar$ units. 
The total Hamiltonian $\hat H (t)$ is given by \cite{MacQuarrie_Mechanical_2013,stern_room-temperature_2022,PhysRevLett.105.140502,Xie_2017,Leggett_1987_Dynamics,Chilingaryan_2015}:
\begin{equation}
\begin{split}
\hat H (t) = & \hat H_S \\
& + \underbrace{ \omega_a \Ad \A + g_a \hat H_a(\Ad + \A) }_{\text{HO$a$ Hamiltonian + HO$a$-spin coupling }} \\
& + \underbrace{\omega_b \Bd \B + g_b \hat H_b(\Bd + \B)}_{\text{HO$b$ Hamiltonian + HO$b$-spin coupling }} \\
&  + \hat H_c (t), 
\end{split}
\label{eq:full_hamiltonian}
\end{equation}
with,
\begin{equation}
\begin{split}
&\hat H_S =  D \hat S_z^2 + \frac{\omega_z}{2} \hat S_z, \\
&\hat H_a =  \cos(\theta_a)\sin(\phi_a) \hat S_x + \sin(\theta_a)\sin(\phi_a) \hat S_y + \cos(\phi_a) \hat S_z, \\
&\hat H_b =  \alpha \left(\cos(\theta_b)\sin(\phi_b) \hat S_x + \sin(\theta_b)\sin(\phi_b) \hat S_y + \cos(\phi_b) \hat S_z \right) \\
& + (1-\alpha) \left( \cos(\gamma_b) \left(\hat S_x^2 -  \hat S_y^2  \right) + \sin(\gamma_b) \left(\hat S_x \hat S_y + \hat S_y \hat S_x  \right)\right), \\
&\hat H_c (t) = ~\vec \Omega (t). \vec S ~\equiv ~\Omega_x(t) \hat S_x + \Omega_y(t)  \hat S_y + \Omega_z(t) \hat S_z.
\end{split}
\end{equation}
The operators $\hat S_x, \hat S_y, \hat S_z$ are spin-1 operators, $\Ad, \A$, and $\Bd, \B$ are respectively creation and annihilation operators of $a$ and $b$. The parameters $\omega_a$ and $\omega_b$ are $a$ and $b$ harmonic oscillator frequencies, whilst $\omega_z$ and $D$ are fixing the spectrum of $S$. For NV-centers, these latter two parameters correspond respectively to the Zeeman effect frequency, and $D$ is a zero field splitting induced by the interaction with the environement~ \cite{MacQuarrie_Mechanical_2013,stern_room-temperature_2022}. $g_a$ and $g_b$ are the coupling constants, and $\theta_i \in [0,2\pi[$, $\phi_i \in [0,\pi]$, and $\gamma_b \in [0,2\pi]$ give us the orientation of the oscillators with respect to the three-level system. The value $\alpha = 0$ or $1$ selects a type of coupling for the second oscillator. Its physical significance is discussed below. The vector $\vec \Omega(t) = (\Omega_x(t),\Omega_y(t),\Omega_z(t)) \in \setR^3$ is the time-dependent control fields. The case without control is simply given by $\vec \Omega (t) = \vec 0 ~\forall t$. Additional constraints (e.g., $\Omega_x (t)=0 $) on the control field may be imposed to avoid an alignment of this latter with the oscillators. In the following, we always consider that $\omega_a$ and $\omega_b$ are fixed parameters, and the spectrum of $S$ can be modified to select a given resonance of the system (for example by tuning the Zeeman splitting $\omega_z$ with a magnetic field). 

\begin{figure}[ht]
	\includegraphics[width=\columnwidth]{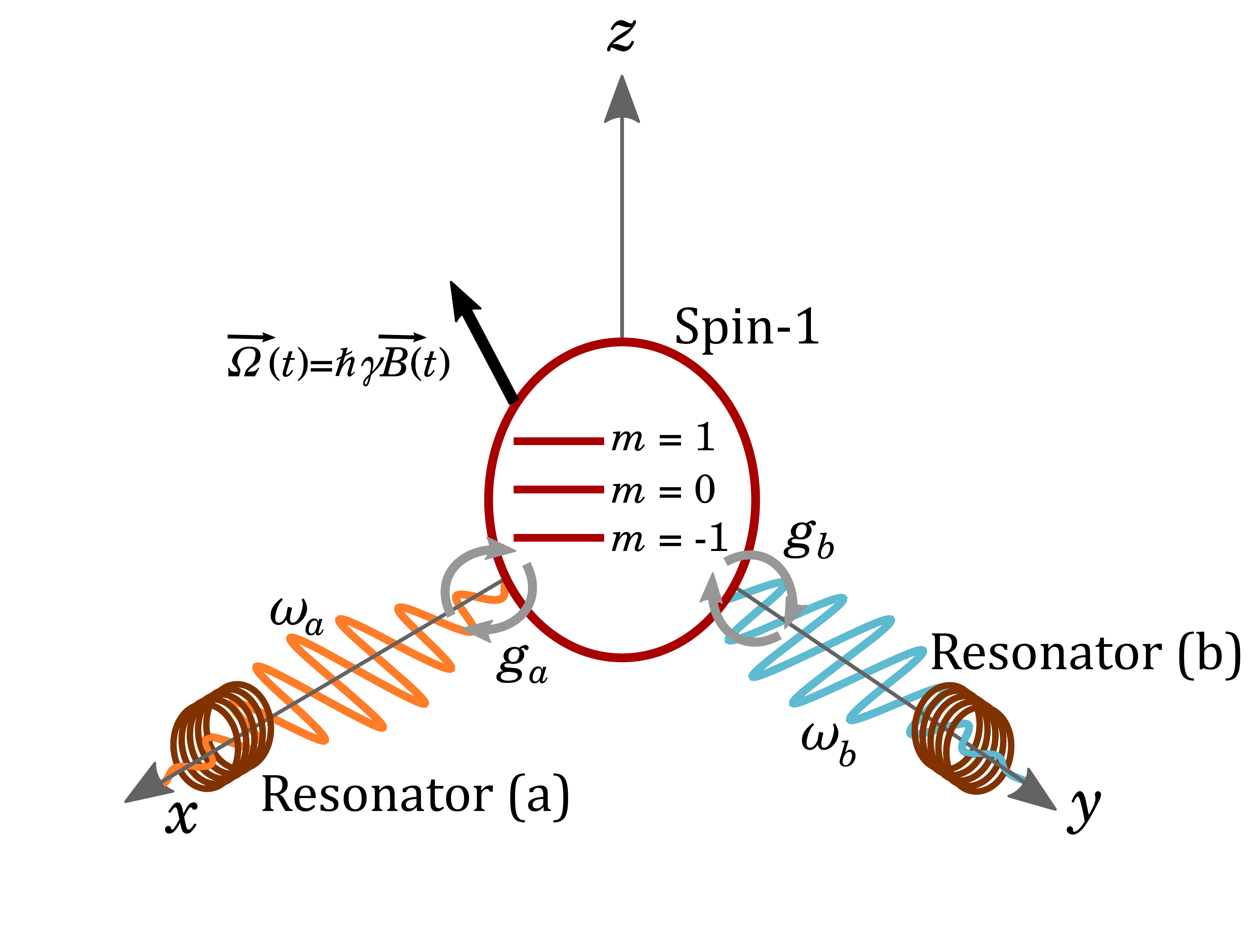}
	\caption{Schematic view of the physical system in the case of $\alpha=1$. The three-level system is represented by a circle at the center of its reference frame. The three internal energy levels are indicated by horizontal lines inside the circle. Quantized oscillators are represented by colored wave-packets. Their related resonators are depicted with small coils. The oscillator $a$ has a wave vector aligned with $x$, and it interacts with the spin on the axis $y$ ($\theta_a = \pi/2$ and $\phi_a = \pi/2$). Similarly, the oscillator $b$ has a wave vector aligned with $y$, and it interacts with the spin on the axis $x$ ($\theta_a =0$ and $\phi_a = \pi/2$). In the general setting, the orientation of the three-level system with respect to the two oscillators and the control field can be adjusted in order to maximize the transfer of quantum excitation between the oscillators. The control field $\vec \Omega (t)$, which is used to enhance the transfer protocol, is depicted by a black arrow (a modulated classical magnetic field).}
	\label{fig:model_system}
\end{figure}

A schematic view of the system is depicted in Fig.~\ref{fig:model_system} in the case $\alpha = 1$. This figure illustrates how the orientation of the inductive coupling between spins and resonators is fixed by the experimental geometry. Contrarily to other parameters, like Zeeman and microwave fields, these parameters cannot be changed easily in situ. Yet our simulations show that having the correct coupling geometry is important to take advantage of the multiple resonances which can appear in a triplet state giving efficient protocols for two-photon exchange, this highlights the importance of simulations to determine the optimal
coupling geometry.

Note that the parameterizations of $\hat H_a$ and $\hat H_b$ are chosen such that they are both diagonalized into $\hat S_z$, and thus, we can always interpret them as a spin operator rotated into a given frame, exactly like the usual Rabi-Hamiltonian of a spin-$\tfrac{1}{2}$ coupled with a harmonic oscillator.  

For later convenience, we also introduce the notation: 
\begin{equation}
\hat H_0  = \hat H_S + \omega_a \Ad \A + \omega_b \Bd \B.
\label{eq:def_H0}
\end{equation}

The broad parameterization of the interaction Hamiltonians allows us to consider several experimental situations~\cite{MacQuarrie_Mechanical_2013,Teissier_strain_2014,awschalom_quantum_2018}. When $\alpha = 0$, the system can model the interaction between the mode of a cavity (oscillator $a$), three energy levels of an NV center,  and a vibration mode $b$ of the system which changes the local spin strain environment and thus its fine structure tensor. In the following, this coupling is called ``\textit{photon-phonon coupling}''. When $\alpha = 1$, we have the more common situation where two cavity modes (not necessarily aligned in the same direction) interact with three energy levels of an atomic system. Thus, this coupling is called ``\textit{photon-photon coupling}''.

We need to quantify the efficiency of a transfer protocol. For this purpose, we use the fidelity with respect to a given target state. In the following, we consider two types of fidelity functions. The first one gives us information on the eigenvectors:
\begin{equation}
F_{eig}[\psi_{target}] = \max_{n} \vert \braket{\psi_{target}}{\phi_n} \vert^2,
\label{def:F_eig}
\end{equation}
with $\left\lbrace \phi_n \right\rbrace_{n=1...\text{dim}\hat H}$ the eigenvectors of $\hat H$. The second function is the fidelity with respect to a target state $\psi_{target}$ at a given time $t$:
\begin{equation}
F[\psi_{target},\psi_0,t,\{\lambda\},\vec \Omega] = \vert \bra{\psi_{target}} \hat U [t,\{\lambda\},\vec \Omega]\ket{\psi_0} \vert^2.
\label{def:F_time}
\end{equation}
Here, $\hat U$ is the evolution operator, $\psi_0$ is the initial state that depends on the transfer protocol, and $\{\lambda\}$ is a shorthand notation to denote the parameters of the Hamiltonian. In practice, we write only the parameters which are relevant to the given context. Several initial and target states are considered in the paper. Their definitions are given below in the text.

\subsection{Interaction with the Environment}

The interaction of quantum systems with their environments leads to numerous decoherence effects that must be taken into account in real-life experiments~\cite{BookBreuer2007}. However, in this paper, we do not study such phenomena. On the basis of recent results \cite{Altafini_2004,Ansel_exploring_2022,Ansel_Optimal_2022}, we can reasonably assume that, for our system, no notable reduction of the relaxation effects can be achieved with coherent controls, as soon as the interaction with the environment is Markovian. A non-Markovian interaction may provide a slight recovery of the system controllability in the presence of dissipation \cite{wilhelm2009,reich_exploiting_2015}, but this is an additional effect that does not directly influence the ones of interest in this paper. However, taking the environment into account is very important for the precise understanding of a particular experimental setup, as in Ref.~\cite{basilewitsch_engineering_2021}. The controllability issue without dissipation is considered in appendix~\ref{sec:controllability}.

A second point to be discussed is the importance of the explicit dependence of the state of $S$ in the definition of $\psi_{target}$, so that we always work in the complete Hilbert space $\Mc H$. In principle, we can trace out the degrees of freedom of $S$, to work with a reduced density matrix for the oscillators $a$ and $b$, and then, \eqref{def:F_eig} and \eqref{def:F_time} can be modified using the quantum fidelity of density matrices~\cite{yuan_fidelity_2017}. This approach using partial traces has gained increasing popularity in the last years with the development of open quantum systems~\cite{BookBreuer2007}. However, in our case, the spin is directly addressed by the control fields, and tracing it out will only complicate the analysis. Moreover, the knowledge of the full state of the system is also necessary to apply an open-loop control strategy in this setting.

\subsection{Resonances and Coupling Regimes}
\label{sec:physical effects}

The three-level system offers multiple resonances that enhance the transfer of one or two excitations. Resonances can be easily found with the degeneracies of $\hat H_0$, defined in Eq.~\eqref{eq:def_H0}. In this paper, two resonances noted (R1) and (R2) are considered. They are defined as follows:
\begin{itemize}
\item \textbf{(R1):} $D = \omega_a - \tfrac{\omega_b}{2}$, $\omega_z = \omega_b$. This resonance produces efficient 1-excitation exchanges. In particular we have potential Rabi-oscillations between the states $\ket{0_S,1_a,0_b}$, $\ket{-1_S,0_a,1_b}$, $\ket{1_S,0_a,0_b}$. They are degenerate eigenvectors of $\hat H_0$ with energy $\omega_a$.
\item \textbf{(R2):} $D = \tfrac{3}{2}(\omega_a - \omega_b)$, $\omega_z = \omega_a - \omega_b$. This resonance produces efficient 2-excitation exchanges. In particular we have potential Rabi-oscillations between the states $\ket{0_S,2_a,0_b}$, $\ket{-1_S,1_a,1_b}$, $\ket{1_S,0_a,2_b}$. They are degenerate eigenvectors of $\hat H_0$ with energy $2\omega_a$.
\end{itemize}

\begin{figure*}[t]
\includegraphics[width=\textwidth]{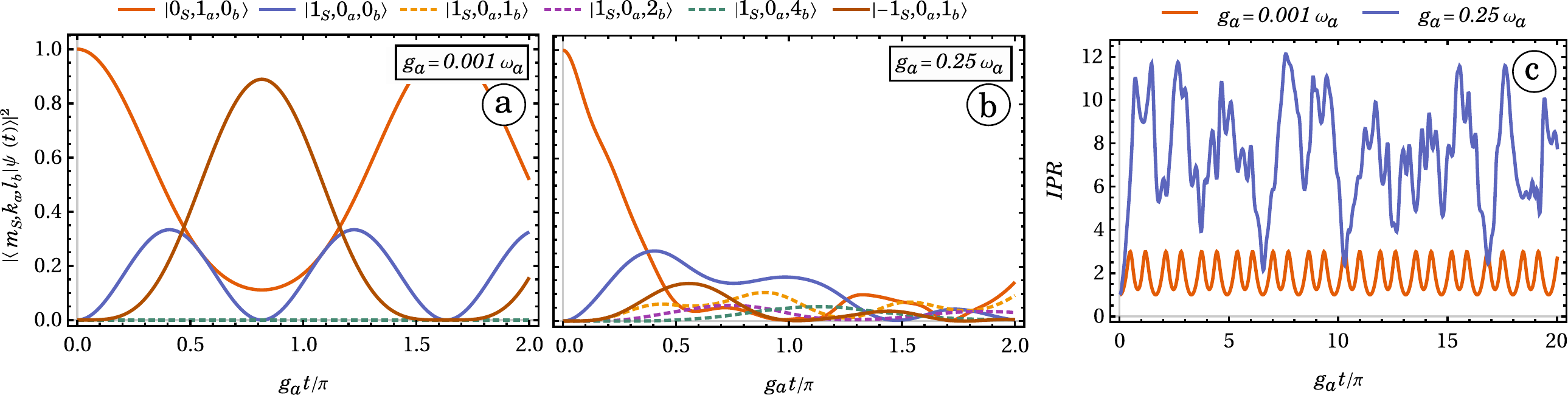}
\caption{\textbf{Panel a) and b)} Population of a few states as a function of time for the initial state $\ket{0_S,1_a,0_b}$. Dynamics in a) and b) are respectively for $g_a = 0.001 \omega_a$ (weak coupling regime) and $g_a = 0.25 \omega_a$ (strong coupling regime, also called ultra-strong). \textbf{Panel c)} Shows the IPR as a function of time for the two dynamics presented in the other panels. In all cases, computations are made with the resonance (R1), and the following parameters: $\omega_b = 0.3 ~\omega_a, D = \omega_a - \omega_b/2, \omega_z=\omega_b, g_a=g_b, \alpha = 0 ,\phi_a = \pi/2, \theta_a=0, \gamma_b=0$.}
\label{fig:dynamics_and_IPR}
\end{figure*}

In addition to the different resonances, the weak and strong coupling regimes (respectively given by $\max(g_a,g_b) \ll \min(\omega_a ,\omega_b)$ and $(g_a,g_b) \sim (\omega_a ,\omega_b)$ ) play a role in the transfer of quantum excitations (the strong coupling can be also qualified of " ultra-strong" or "deep-strong" when the physical system has other characteristic frequencies).
The main difference between the regime is the projection of the eigenstates onto a few (weak coupling) or many (strong coupling) states of the canonical basis $\{\ket{m_S,k_a,l_b}\}$. In the second case, if the initial state is $\ket{m_S,k_a,l_b}$, we observe a quick diffusion of the quantum excitation and a loss of fidelity. This effect can be illustrated with plots of dynamics of the populations (see Fig.~\ref{fig:dynamics_and_IPR} a) and b)), or by computing the Inverse Participation Ratio (IPR)~\cite{Kramer_1993} of the eigenstates. The IPR of a pure state $\ket{\psi}=\sum_n \psi_n \ket{n}$ is given by: 
\begin{equation}
IPR= \left(\sum_n |\psi_n|^4 \right)^{-1}.
\end{equation}
Roughly, it gives us the number of basis states on which $\ket{\psi}$ is decomposed with equal weights. In Fig.~\ref{fig:dynamics_and_IPR}, examples of state populations and the IPR are plotted as a function of time for different coupling strengths and for an initial state $\ket{0_S,1_a,0_b}$. In the weak coupling regime ($g_a = 0.001 ~ \omega_a$, see panel a)), the IPR (panel c)) fluctuates between 1 and 3, confirming that the quantum states in panel a) is always a superposition of $\ket{0_S,1_a,0_b}$, $\ket{-1_S,0_a,1_b}$, and $\ket{1_S,0_a,0_b}$. The quantum excitation, initially in the oscillator $a$, is partially transferred to the oscillator $b$, and then, it comes back to the oscillator $a$. In the strong coupling regime ($g_a = 0.25 ~ \omega_a$, see the panel b)), the quick dispersion of the excitation onto many excited states reduces considerably the transfer, and no periodicity is observed. The IPR is mostly between 5 and 12 (with sometimes smaller or larger values), but it never goes back to 1 (at least, not at a reasonable time).
The dispersion effect can also be shown with the diagonalization of the Rabi Hamiltonian. Explicit calculations are made in appendix~\ref{sec:rabi}. 

\begin{figure*}[t]
\includegraphics[width=\textwidth]{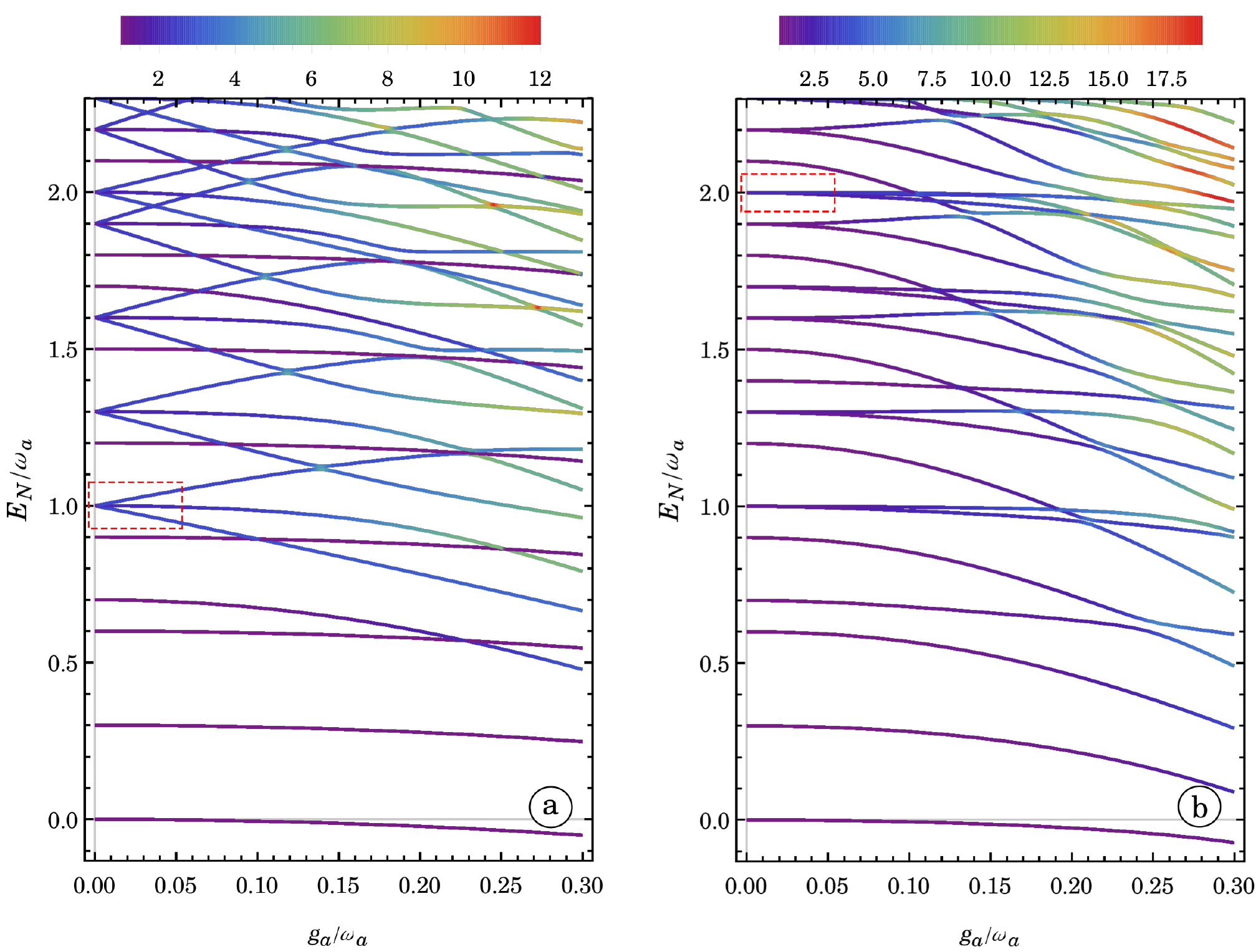}
\caption{
\textbf{Panel a)} Energy spectrum of the system in the case of (R1) as a function of the coupling constant $g_a$. The color map shows the IPR of the corresponding eigenstate. Other parameters are: $\omega_b = 0.3 ~\omega_a, g_b= g_a/\sqrt{2}, \alpha = 0 ,\phi_a = \pi/2, \theta_a=0, \gamma_b=0$. \textbf{Panel b)} same as the panel a) but with the resonance (R2) and the parameters: $\omega_b = 0.3 ~\omega_a, g_b=g_a/2, \alpha = 1 ,\phi_a = 0.210573 ~\pi, \theta_a=0, \phi_b = 0.243693~ \pi, \theta_b=0$. In the two panels, the areas of interest in the weak coupling regime are emphasized with dashed red rectangles ($E_N \approx \omega_a$ and $E_N \approx 2\omega_a$). The values of the parameters used to construct the spectrum are motivated by the results presented in Sec.~\ref{sec:weak_coupling}.}
\label{fig:spectrum}
\end{figure*}

In this paper, we focus on exchanges near the ground state of each oscillator, but this is not strictly necessary. To provide a wide view of the different resonance mechanisms at different energy levels, we show in Fig.~\ref{fig:spectrum} two examples of energy spectrum as a function of the coupling strength. In order to underline in which conditions we can observe the dispersion effect, we show with a color code the IPR of the eigenvectors. Interestingly, there are eigenstates where the IPR remains equal to 1 and they are not affected by the dispersion. The strong coupling regime starts around $g_a = 0.1 \omega_a$, where we have a crossing between the energy levels of interest (both for (R1) and (R2)). The dispersion effect becomes very strong around $g_a = 0.2 \omega_a$, due to a noticeable increase in the IPR of most of the eigenvectors.

We close this section with a comment on the experimental relevance of the weak and strong coupling regimes for the two couplings. For both photon-photon and photon-phonon couplings, the weak coupling regime is well justified and they are routinely considered in experiments \cite{Lee_2017}. The strong coupling regime is harder to reach but several techniques can be used, such as artificial atoms \cite{PhysRevLett.105.237001} or collective ensembles \cite{kurizki_quantum_2015,PhysRevLett.105.140502}. As far as we know, there is not yet an experimental setup where NV-center spins are strongly coupled to phonons, but a strong phonon-atoms coupling has been realized in \cite{PhysRevLett.117.103201}. 

\section{Quantum Exchanges in the Weak Coupling Regime}
\label{sec:weak_coupling}

We consider in this section the transfer of a quantum excitation between the two oscillators in the weak coupling regime. We start the analysis with the resonance (R1), and we finish with (R2). In both cases, the system is not controlled (i.e., $\vec \Omega (t) = \vec 0 ~ \forall t$). The cases with control fields are studied in the next sections.

\subsection{1-excitation exchange}
\label{sec:weak_case1}

We consider the transfer of a single quantum excitation (first excited oscillator state) between the two 
oscillators using the resonance (R1). To produce the exchange, we need Rabi-oscillations between the states $\ket{0_S,1_a,0_b}$ and $\ket{-1_S,0_a,1_b}$. A simple way to obtain such oscillations is to determine under which conditions $\ket{\psi_{target}}= \tfrac{1}{\sqrt{2}} \left(\ket{0_S,1_a,0_b}+ \ket{-1_S,0_a,1_b}\right)$ is an eigenvector of the Hamiltonian. For $g_a$ and $g_b$ sufficiently small, $F_{eig}(\psi_{target}) \approx 1$ can be obtained using a first-order perturbative expansion of the Hamiltonian. 
At this order of perturbation, the perturbed Hamiltonian restricted to the subspace defined by $\ket{0_S,1_a,0_b}$, $\ket{-1_S,0_a,1_b}$, and $\ket{1_S,0_a,0_b}$ is simply given by $\hat H_{pert.~R1} = \textstyle \sum_{i,j} \bra{i} \hat H \ket{j}  \ket{i} \bra{j}$, where $i,j$ denote the three possible states of the subspace. A straightforward calculation gives us:
\begin{equation}
\begin{split}
&\hat H_{pert.~R1} = \\
& ~~ ~ + \frac{g_a}{\sqrt{2}}\sin(\phi_a)(\cos(\theta_a)-\ii\sin(\theta_a)) \ket{0,1,0}\bra{1,0,0} \\
& ~~ ~ + g_b(1-\alpha)(\cos(\gamma_b)-\ii \sin(\gamma_b)) \ket{-1,0,1}\bra{1,0,0}\\
& ~~ ~ +H.C.
\end{split}
\label{eq:Heff_1}
\end{equation}

Note that for simplicity, the subscripts $S, a, b$ of each quantum number, and an overall constant term $\omega_a$ in the Hamiltonian have been omitted. Approximated eigenvectors are determined by diagonalizing $\hat H_{pert.~R1}$. Three eigenvectors are obtained, but only one of them can give us a solution. Without normalization, this latter is
\begin{equation*}
\ket{-1_S,0_a,1_b} - \frac{g_b}{g_a}\sqrt{2}(1-\alpha) e^{-\ii (\gamma_b - \theta_a)} \text{csc}(\phi_a) \ket{0_S,1_a,0_b}.
\end{equation*}
Then, the target state $\ket{\psi_{target}}$ can be obtained  only if $\alpha = 0$, $\gamma_b - \theta_a = k \pi,~k\in\setZ$,  and $ \sin(\phi_a)= \pm \sqrt{2}g_b/g_a$. We recall that $\alpha = 0$ corresponds to a photon-phonon coupling. The last constraint gives us a necessary condition, which is $g_b/g_a \leq \sqrt{2}$.

Despite the condition $\alpha=0$, solutions with $\alpha=1$ (photon-photon coupling) can be obtained with a slight modification of the problem. A first possibility is to consider a second-order expansion, to lift the residual degeneracy of the system. An example of second-order expansion is considered in Sec.~\ref{sec:weak_case2}. Another possibility is to keep the first order of perturbation but to modify the resonance. For example, using $D= \tfrac{1}{2}(\omega_a+\omega_b)$ and $\omega_z = \omega_b - \omega_a $, we have a perturbed subspace given by the states $\ket{1_S,1_a,0_b}$, $\ket{0_S,1_a,1_b}$, and $\ket{-1_S,0_a,1_b}$. Similar calculations as the ones presented here-above can give us a simple condition where $\tfrac{1}{\sqrt{2}} (\ket{1_S,1_a,0_b} + \ket{-1_S,0_a,1_b})$ is an eigenvector at the first order of perturbation.

\subsection{2-excitation exchange}
\label{sec:weak_case2}

\begin{figure}[t]
\includegraphics[width=\columnwidth]{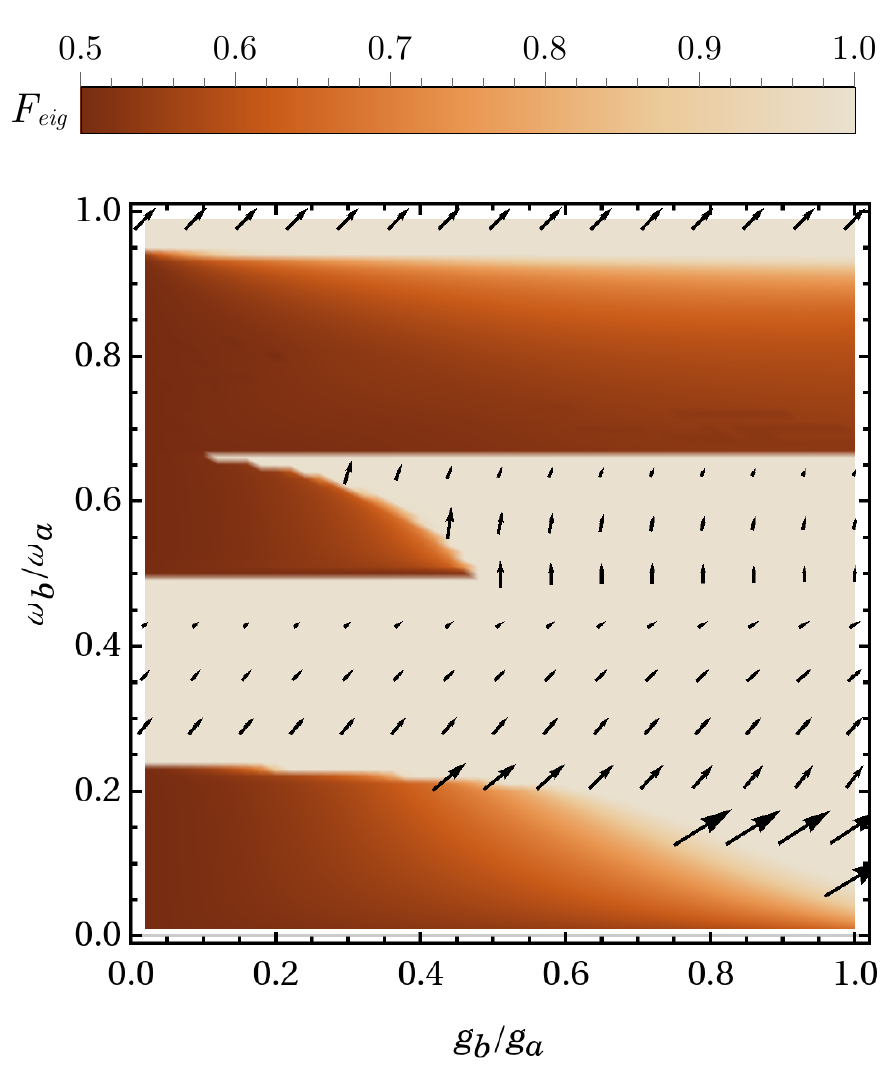}
\caption{The density plot shows the maximum of the fidelity function $F_{eig}$ with $\ket{\psi_{target}} = \tfrac{1}{\sqrt{2}}\left( \ket{0_S,2_a,0_b} + \ket{1_S,0_a,2_b}\right)$, computed in the weak coupling limit of the resonance (R2), for different values of $\omega_b/\omega_a$ and $g_b/g_a$. At each point of the graph, the parameters $\phi_a$ and $\phi_b$ are optimized. Calculations are made using the perturbed Hamiltonian~\eqref{eq:Heff_2}, such that the results do not depend on $g_a/\omega_a$. For several points with $F_{eig}>0.999$, a small black vector gives the optimal solution with the smallest value of $\phi_a$. The abscissa and the ordinate of the vector give respectively $\phi_a$ and $\phi_b$. The scale is $0.097:\pi$. Other physical parameters are : 
$\alpha = 1$ (photon-photon coupling), $\theta_a = \theta_b =0$.}
\label{fig:R2_WC_1}
\end{figure}
\begin{figure}[t]
\includegraphics[width=\columnwidth]{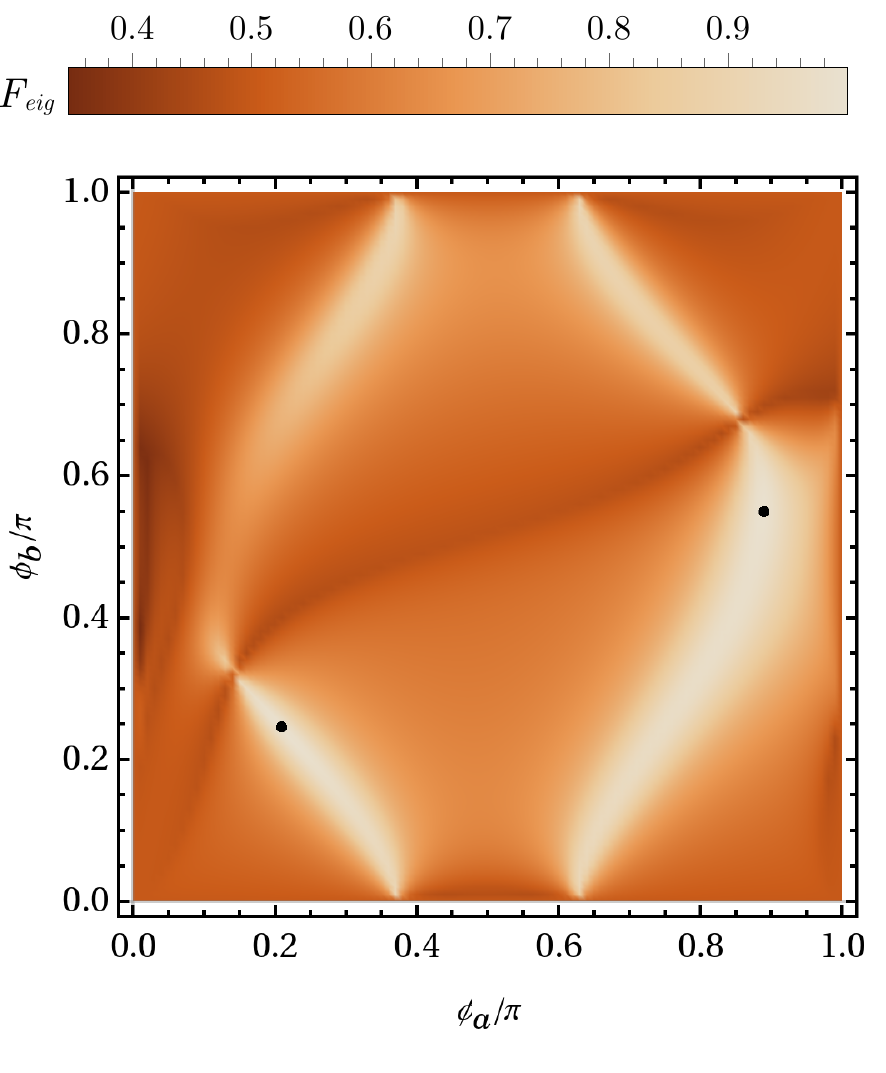}
\caption{Fidelity $F_{eig}$ with the target state $\ket{\psi_{target}} = \tfrac{1}{\sqrt{2}}\left( \ket{0_S,2_a,0_b} + \ket{1_S,0_a,2_b}\right)$ computed in the weak coupling limit of the resonance(R2), and in the case of $\omega_b/\omega_a = 0.3$ and $g_b/g_a = 0.5$. The two black dots indicate the location of the global maximums ($F_{eig}=1$). Note that the fidelity is $\pi$-periodic. Other physical parameters are : 
$\alpha = 1$ (photon-photon coupling), $\theta_a = \theta_b =0$.}
\label{fig:R2_WC_2}
\end{figure}

The study of two excitation exchanges using (R2) is more difficult than the case of Sec.~\ref{sec:weak_case1}. To simplify the analysis, we assume $\theta_a = \theta_b =0$ and $\alpha = 1$ (photon-photon coupling). Similar results can be obtained if we relax these assumptions. Then, we rewrite $\hat H_a$ and $\hat H_b$ in a simpler form:
\begin{equation}
\begin{split}
\hat H_a & = \sin \phi_a \hat S_x + \cos \phi_a \hat S_z \\
\hat H_b & = \sin \phi_b \hat S_x + \cos \phi_b \hat S_z.
\end{split}
\label{eq:Ha_and Hb_2_case2}
\end{equation}
Following the same idea as in Sec.~\ref{sec:weak_case1}, we determine the values of $\phi_a$ and $\phi_b$ that maximize $F_{eig}$, for which the target state is $\ket{\psi_{target}} = \tfrac{1}{\sqrt{2}} \left( \ket{0_S,2_a,0_b} + \ket{1_S,0_a,2_b}\right)$. 

With the resonance (R2), an effective Hamiltonian is constructed similarly to the one of Sec.~\ref{sec:weak_case1}, but using the next perturbation order. We define 
\begin{equation}
\hat H_{pert.~R2} = \textstyle \sum_{i,j,m} \tfrac{\bra{i} \hat H_a+\hat H_b \ket{m} \bra{m} \hat H_a+\hat H_b \ket{j} }{\bra{i} \hat H_0 \ket{i} - \bra{m}\hat H_0 \ket{m}} \ket{i} \bra{j},
\end{equation}
where $i,j$ denotes the three states $\ket{0_S,2_a,0_b}$, $\ket{-1_S,1_a,1_b}$, $\ket{1_S,0_a,2_b}$, and $m\neq i,j$ is a state outside the degenerated subspace. Note that in the denominator, the difference of energy is between the unperturbed energy of the subspace $\bra{i} \hat H_0 \ket{i}$ and the unperturbed energy of states outside the subspace $\bra{m}\hat H_0 \ket{m}$. A straightforward calculation gives us:
\begin{equation}
\begin{split}
& \hat H_{pert.~R2} = \\
&  \left(\begin{array}{ccc}
V_{\ket{-1,1,1}\bra{-1,1,1}} & V_{\ket{0,2,0}\bra{-1,1,1}}  & V_{\ket{1,0,2}\bra{-1,1,1}}  \\ 
V_{\ket{0,2,0}\bra{-1,1,1}}  & V_{\ket{0,2,0}\bra{0,2,0}}  & 0 \\ 
V_{\ket{1,0,2}\bra{-1,1,1}}  & 0 & V_{\ket{1,0,2}\bra{1,0,2}} 
\end{array}  \right),
\end{split}
\label{eq:Heff_2}
\end{equation}
with,
\begin{align*}
V_{\ket{-1,1,1}\bra{-1,1,1}} &= \frac{1}{2} \left(\frac{g_a^2 \sin ^2(\phi_a) (3 \omega_b-4 \omega_a)}{\omega_b(2 \omega_a-\omega_b)}-\frac{2 g_a^2 \cos ^2(\phi_a)}{\omega_a} \right. \\
& \left.+\frac{g_b^2 \sin ^2(\phi_b) (3 \omega_a-2 \omega_b)}{\omega_a (\omega_a-2 \omega_b)}-\frac{2 g_b^2 \cos ^2(\phi_b)}{\omega_b}\right),\\
V_{\ket{0,2,0}\bra{-1,1,1}} &= g_a g_b \left(\frac{\cos (\phi_a) \sin (\phi_b)}{\omega_a}-\frac{\sin (\phi_a) \cos (\phi_b)}{\omega_b}\right),\\
V_{\ket{1,0,2}\bra{-1,1,1}} &= \frac{3 g_a g_b \sin (\phi_a) \sin (\phi_b) (\omega_a-\omega_b)}{\sqrt{2} (\omega_a-2 \omega_b) (2 \omega_a-\omega_b)} \\
V_{\ket{0,2,0}\bra{0,2,0}} &=  \frac{1}{2} \left(g_a^2 \sin ^2(\phi_a) \left(\frac{3}{\omega_b-2 \omega_a} \right. \right. \\
& \left. +\frac{3}{2 \omega_b-3 \omega_a}-\frac{2}{\omega_a-2 \omega_b}+\frac{2}{\omega_b}\right) \\
& \left. +g_b^2 \sin ^2(\phi_b) \left(\frac{1}{\omega_b-2 \omega_a}-\frac{1}{\omega_a}\right)\right), \\
\end{align*}
\begin{align*} 
V_{\ket{1,0,2}\bra{1,0,2}} &= \frac{g_a^2 \sin ^2(\phi_a)}{2 \omega_a-4 \omega_b}-\frac{g_a^2 \cos ^2(\phi_a)}{\omega_a} \\
& +\frac{g_b^2 \sin ^2(\phi_b) (10 \omega_a-9 \omega_b)}{8 \omega_a^2-16 \omega_a \omega_b+6 \omega_b^2}-\frac{g_b^2 \cos ^2(\phi_b)}{\omega_b}.
\end{align*}
For clarity, we have omitted the index $S,a,b$ of each quantum number and an overall constant term $2\omega_a$ in the Hamiltonian. 
We note that $\omega_b/\omega_a = 0,\tfrac{1}{2}, \tfrac{2}{3},2$ are singular values for which no solution can be found at this order of perturbation. 

Due to the relative complexity of the coefficients $V_{ij}$ in Eq.~\eqref{eq:Heff_2}, the existence of a solution is not guaranteed for any possible value of the system parameters, and the calculation of $\phi_a$ and $\phi_b$ is not a simple task. Additionally, several nonequivalent solutions can exist. However, the problem can be solved rather easily with the numerical calculation of $\max_{\phi_a,\phi_b} \left( F_{eig} \right)$. We refer to the Appendix~\ref{sec:numerical_methods} for details on the numerical optimizations.
We report in Fig.~\ref{fig:R2_WC_1} the maximum of $F_{eig}$ for the parameters $\omega_b/\omega_a \in ]0,1[\setminus 0.5$ and $g_b/g_a \in[0,1]$. We observe that the space of parameters is divided into several areas, with $F_{eig} =1$ or with $F_{eig} < 1$. This corresponds to areas where a solution exists or to areas without a solution.
We also observe that an optimal solution depends strongly on the system parameters. In particular, we note that solutions are rather robust against variations of $g_a/g_b$, and on the opposite, they are not robust against variations of $\omega_b/\omega_a$. However, this is not a big issue since the experimental stability of cavity modes is in general very good. 
 
In order to illustrate the existence of different equivalent solutions, we plot in Fig.~\ref{fig:R2_WC_2} the fidelity $F_{eig}$ as a function of $\phi_a$ and $\phi_b$, in the case of $\omega_b/\omega_a = 0.3$ and $g_b/g_a=0.5$. Note the existence of four suboptimal maximums of the fidelity, that must be avoided in the numerical optimization.

Finally, we underline that with this kind of resonance involving three states, we obtain a faster transition than with a resonance that couples only the two states of interest. Indeed, when the parameters of the Hamiltonian are fixed to provide a simple resonance between $\ket{ 0_S, 2_a, 0_b }$ and $\ket{ 1_S, 0_a, 2_b}$, the couplings to the spin also lift this degeneracy. This leads to two resonant eigenstates $\ket{0_S, 2_a, 0_b} \pm \ket{ 1_S, 0_a, 2_b}$ one of which is $\psi_{target}$.
However, the interaction matrix element between the two uncoupled states is, in this case, given by a fourth-order process in the strength of the coupling between spin and oscillators with an amplitude 
proportional to $g_a^2 g_b^2$. As a result, the simple resonance leads to very small energy splitting for small couplings $g_a, g_b$. The more stringent three-state-resonance condition we investigated allows us to obtain coupling matrix elements that are only of second order in $g_a$ and $g_b$ (see Eq.~\eqref{eq:Heff_2}). The energy splitting of the target state with the other eigenstates of the Hamiltonian limits ultimately the rate at which this state can be prepared. Thus, the three-state-resonance condition is much more efficient in the weak coupling regime.

\section{Effective Hamiltonian Using a Time-Dependent Control Field}
\label{sec:effective_hamiltonian}

In Sec.~\ref{sec:weak_coupling}, we have derived conditions for which one eigenvector of the Hamiltonian gives us an efficient transfer of quantum excitation (in the perturbative approximation). However, we have seen that for a given set of values for $(\omega_a, \omega_b, g_a, g_b)$, a solution may not always exist, or it can be experimentally too difficult to obtain the optimal angles. In this section, we provide a method that enlarges, to some extent, the possibilities of Hamiltonian engineering when the method described previously fails. 

The general idea is to employ a parameterized control field $\vec \Omega(t)$ that produces an effective Hamiltonian, which depends on the control parameters. Then, we can proceed similarly as in Sec.~\ref{sec:weak_coupling}, and we adjust the parameters of the control to maximize $F_{eig}.$
The effective Hamiltonian is derived using averaged Hamiltonian theory~\cite{maricq_application_1982,brinkmann_introduction_2016}. Similar methods have already been applied in many contexts, such as (to cite a few) in magnetic resonance \cite{maricq_application_1982}, optical lattices \cite{PhysRevA.90.031601} or to study non-Markovian dynamics \cite{janowicz_non-markovian_2000}.

We illustrate the method with a specific example in the case of the resonance (R1) for which the final result gives us a simple solution. The same method can be applied to (R2) or more complicated scenarios. We choose the parameters of the Hamiltonian such that $\hat H_a = \hat S_x$ and $\hat H_b = \hat S_x^2 - \hat S_y^2$. This corresponds to a photon-phonon coupling. We also assume that that $\Omega_x(t) = 0$, $\Omega_y(t) = A_c \cos (\Omega_c t) \cos(\theta_c)$, and $\Omega_z(t) = A_c \cos (\Omega_c t) \sin(\theta_c)$, where $A_c$, $\Omega_c$, and $\theta_c$ are the parameters of the control field that must be determined. Similar calculations can be performed with the general Hamiltonian of Eq.~\eqref{eq:full_hamiltonian}.

The first step of our effective Hamiltonian construction is to make a change of frame of the evolution operator~\cite{messiah1962quantum}:
\begin{widetext}
\begin{equation}
\begin{split}
\hat U(t) & = \mathbb{T}\exp \left( - \ii \int_0^t dt' \left[ \hat H_0 + g_a \hat S_x (\Ad + \A) + g_b (\hat S_x^2 - \hat S_y^2)(\Bd+\B) + \hat H_c(t') \right] \right) \\
&= U_c (t) \mathbb{T}\exp \left( - \ii \int_0^t dt' U_c ^\dagger (t')\left[ \hat H_0 + g_a \hat S_x (\Ad + \A) + g_b (\hat S_x^2 - \hat S_y^2)(\Bd+\B) \right]U_c(t') \right) \\
&\approx U_c (t) \exp \left( - \ii \int_0^t dt' U_c ^\dagger (t')\left[ \hat H_0 + g_a \hat S_x (\Ad + \A) + g_b (\hat S_x^2 - \hat S_y^2)(\Bd+\B) \right]U_c(t') \right),
\end{split}
\label{eq:approx_evol_op_magnus}
\end{equation}
\end{widetext}
where $\mathbb{T}$ is the time ordering operator, and a first-order Magnus expansion~\cite{Blanes_Magnus_2009} is made in the last line. We have also introduced
\begin{equation}
\begin{split}
&\hat U_c(t) = \mathbb{T}\exp\left(-\ii \int_0^t H_c(t') dt'\right) \\
& = \exp\left(-\ii \frac{A_c}{\Omega_c}\sin(\Omega_c t)\left[\cos(\theta_c) \hat S_y + \sin(\theta_c) \hat S_z\right] \right).
\end{split}
\end{equation}
The error introduced by neglecting second-order terms of the Magnus expansion is discussed in Appendix~\ref{sec:convergence_magnus}.
From Eq.~\eqref{eq:approx_evol_op_magnus}, we can define three effective Hamiltonian operators:
\begin{align}
\hat H_S ' & = \frac{1}{t}\int_0^t dt' U_c^\dagger(t') \hat H_S U_c(t'), \\
\hat H_a ' & = \frac{1}{t}\int_0^t dt' U_c^\dagger(t') \hat S_x U_c(t'), \\
\hat H_b ' & = \frac{1}{t}\int_0^t dt' U_c^\dagger(t') (\hat S_x^2 - \hat S_y^2) U_c(t').
\end{align}
We provide here the explicit calculations of $\hat H_a'$. The explicit formula of $\hat H_S '$ and $\hat H_b '$ are derived by following the same computation steps. A straightforward calculation of $U_c^\dagger(t') \hat S_x U_c(t')$ gives us:
%
\begin{equation}
\begin{split}
\hat H_a' =\frac{1}{t}\int_0^t dt' & \left(  \cos(B) \hat S_x - \sin(\theta_c) \sin(B) \hat S_y  \right.\\
& \left. + \cos(\theta_c) \sin(B) \hat S_z \right),
\end{split}
\end{equation}
with $B = A_c\sin(\Omega_c t')/\Omega_c$. Next, we follow a similar calculation as in Refs.~\cite{janowicz_non-markovian_2000}. We use the Jacobi–Anger expansion~\cite{abramowitz1948handbook}: $e^{\ii d \sin(\Omega_c t)} = \sum_{n=-\infty}^\infty J_n (d) e^{\ii n \Omega_c t}$, with $J_n(d)$ the $n$-th Bessel function of the first kind, to determine:
\begin{equation}
\begin{split}
\frac{1}{t}\int_0^t dt' \cos(B) = & J_0\left(\frac{A_c}{\Omega_c}\right) \\
&+ 2 \sum_{n=1}^\infty J_{2n} \left(\frac{A_c}{\Omega_c}\right) \text{sinc} (2n \Omega_c t), 
\end{split}
\label{eq:int_cosB}
\end{equation}
\begin{equation}
\begin{split}
\frac{1}{t}\int_0^t dt' \sin(B)  = & 2 \sum_{n=1}^\infty J_{2n-1} \left(\frac{A_c}{\Omega_c}\right) \\
& \times \frac{2 \sin^2 ((2n-1) \Omega_c t/2)}{(2n-1)\Omega_c t} .
\end{split}
\label{eq:int_sinB}
\end{equation}
For $\Omega_c$ sufficiently large, all the terms with $n \geq 1$ are negligible in Eq.~\eqref{eq:int_cosB} and Eq.~\eqref{eq:int_sinB}, and the system becomes approximately time independent, i.e., $\frac{1}{t}\int_0^t dt' \cos(B)  \approx J_0\left(\frac{A_c}{\Omega_c}\right)$ and $\frac{1}{t}\int_0^t dt' \sin(B)  \approx 0$. We finally deduce that:
\begin{equation}
\hat H_a' \approx  J_0\left(\frac{A_c}{\Omega_c}\right) \hat S_x.
\label{eq:H_a'}
\end{equation}
Note that the sinusoidal control introduces a scaling factor $J_0(A_c/\Omega_c)$ of the coupling strength, whose absolute value is smaller than 1. Therefore, the effective Hamiltonian cannot leave the weak-coupling regime. By setting $J_0(A_c/\Omega_c)=0$, we can also remove the coupling of the three-level system with the oscillators (see~\cite{janowicz_non-markovian_2000} for an illustration of the effect with another physical system).

The same calculation steps applied to $\hat H_S'$ and $\hat H_b'$ give us approximated expression of the operators, which depends only on the values $A_c/\Omega_c$ and $\theta_c$. The result is very lengthy, but we look ahead with the results and we present their expression with the optimal value $\theta_c = \pi/2$, for which we have a simple result:
\begin{align}
\hat H_S' & \approx \hat H_S, \\
\hat H_b' & \approx J_0\left(2\frac{A_c}{\Omega_c}\right) (\hat S_x^2 - \hat S_y^2).
\label{eq:H_b'}
\end{align}
We see that the sinusoidal control changes the values of the coupling constant, such that $g_a \rightarrow g_a J_0\left(A_c/\Omega_c\right)$ and $g_b \rightarrow g_b J_0\left(2 A_c/\Omega_c\right)$. This result is interesting in the situation of Sec.~\ref{sec:weak_case1}, where $g_a = g_b$. In this configuration, there is no solution for $ \sin(\phi_a)= \pm \sqrt{2}g_b/ g_a$, but we have a solution for $ \sin(\phi_a)= \pm \sqrt{2}J_0\left(2 A_c/\Omega_c\right)J_0\left(A_c/\Omega_c\right)^{-1}$.

\begin{figure}
\includegraphics[width=\columnwidth]{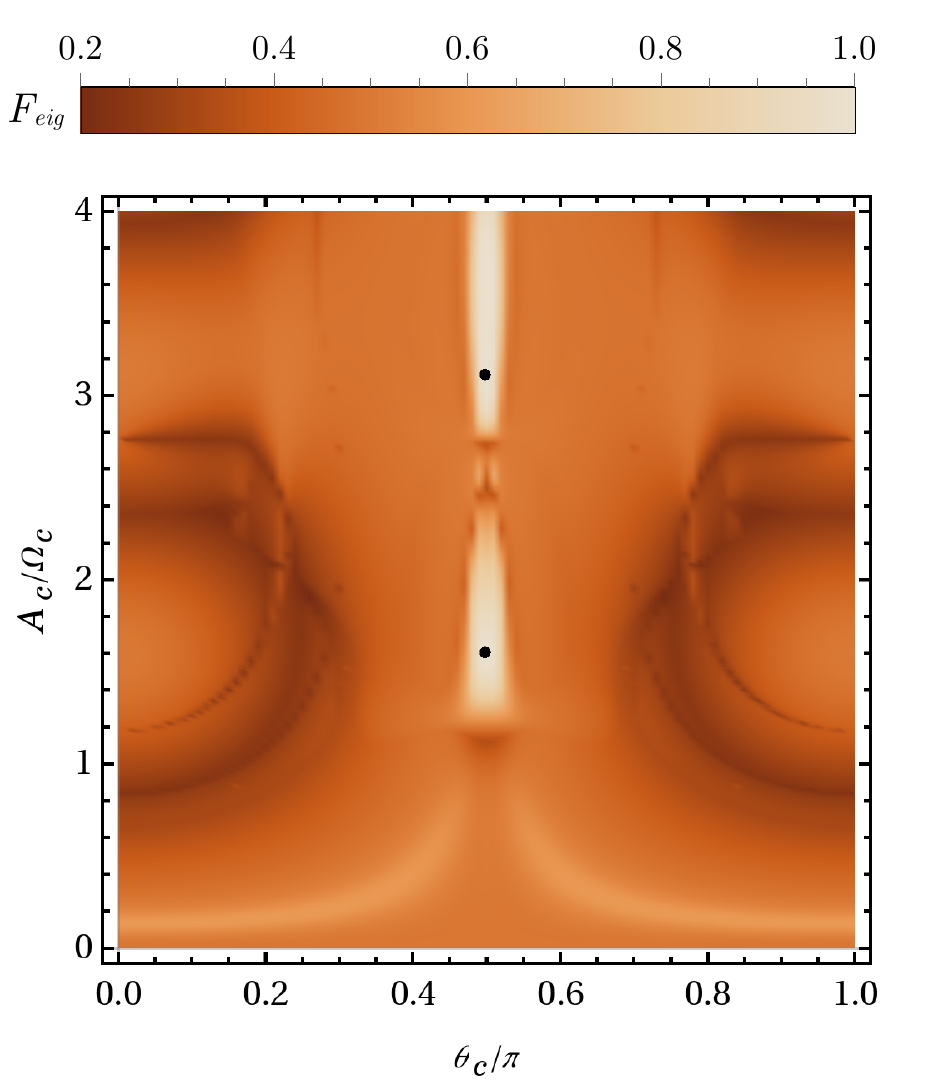}
\caption{Fidelity $F_{eig}$ as a function of the sinusoidal control field parameters, in the case of the resonance (R1), and in the weak coupling regime. The target state is $\ket{\psi_{target}}=\tfrac{1}{\sqrt{2}}\left( \ket{0_S,1_a,0_b}+\ket{-1_S,0_a,1_b}\right)$. The black points show the location of two global maximums ($F_{eig}>0.999$). They have the coordinates: $\theta_c = \pi/2$ and $A_c/\Omega_c \approx 1.60202$ and $3.11897$. The parameters used in the simulations are $\omega_b = 0.3~  \omega_a, D_s = \omega_a - \omega_b/2, \omega_z = \omega_b, g_a = g_b = 0.01 ~ \omega_a,\theta_a=0,\phi_a=\pi/2, \alpha=0, \gamma_b=0, \Omega_c = 3~ \omega_a, A_c = 1.60202~  \Omega_c $.}
\label{fig:sinusoidal_control_1}
\end{figure}
\begin{figure}
\includegraphics[width=\columnwidth]{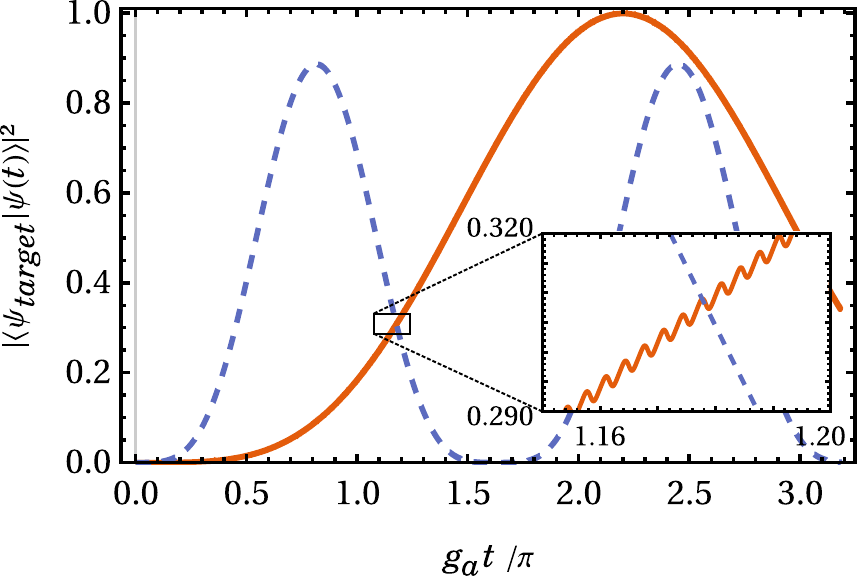}
\caption{Time evolution of $\vert \braket{\psi_{target}}{\psi(t)}\vert^2$ without control field (dashed blue line), and with a sinusoidal control field (solid red line), in the case of the resonance (R1), and in the weak coupling regime. The target state is $\ket{\psi_{target}}=\tfrac{1}{\sqrt{2}}\left( \ket{0_S,1_a,0_b}+\ket{-1_S,0_a,1_b}\right)$. In order to show the quality of the approximation, the simulation is performed using the full quantum system. The maximum of the red curve is $\approx 0.9993$. The inset shows the small fluctuations of the probability which are neglected by the first-order approximation of the  Jacobi-Anger expansion. The parameters used in the simulations are $\omega_b = 0.3~  \omega_a, D_s = \omega_a - \omega_b/2, \omega_z = \omega_b, g_a = g_b = 0.01 ~ \omega_a,\theta_a=0,\phi_a=\pi/2, \alpha=0, \gamma_b=0, \Omega_c = 3~ \omega_a, A_c = 1.60202~  \Omega_c $.  }
\label{fig:sinusoidal_control_2}
\end{figure}

The optimal solution is found with the numerical maximization of the fidelity $F_{eig}$ with respect to $A_c/\Omega_c$ and $\theta_c$. Details on the numerical optimizations are given in Appendix~\ref{sec:numerical_methods}. The density plot of $F_{eig}$  as a function of the two relevant parameters is shown in Fig.~\ref{fig:sinusoidal_control_1}. There are several maximums with fidelity higher than 0.999. The one giving the largest effective coupling is obtained when $\theta_c = \pi/2$ and $A_c/\omega_c \approx 1.60202$. For this solution, we have $J_0(A_c/\Omega_c)\approx 0.4542$, $J_0(2 A_c/\Omega_c)\approx -0.3212$, and  $J_0\left(2 A_c/\Omega_c\right)J_0\left(A_c/\Omega_c\right)^{-1} \approx -0.7071 \approx -1/\sqrt{2}$. This is in agreement with the results obtained in Sec.~\ref{sec:weak_case1}. In order to show the efficiency of the control scheme, we show in Fig.~\ref{fig:sinusoidal_control_2} an example of dynamics in the cases with the optimal sinusoidal control field, and without the control field. To show the quality of the approximation, the simulation is performed using the full quantum system. The maximum obtained with the control field is $\vert \braket{\psi_{target}}{\psi(t)}\vert^2 \approx 0.9993$. The target state is reached with very good fidelity.

In the specific example presented here above, the optimal solution is given by a rescaling of the coupling constant. Nonetheless, for an arbitrary $\theta_c$ the effective operators $H_S'$ and $H_b'$ are given by a complex expression with sums and products of $\hat S_{x,y,z}$ (not shown in this paper due to the size of the formulas). This can be useful to correct the alignment of the oscillators with respect to the three-level system if we have an experimental defect, or if it is experimentally too difficult to obtain the optimal angles given in Sec.~\ref{sec:weak_coupling}. For example, in the situations described in Fig~\ref{fig:R2_WC_1} and \ref{fig:R2_WC_2}, an effective Hamiltonian may provide enough flexibility to reach the optimal configuration if this latter cannot be realized experimentally.

\section{Quantum Exchanges in the Strong Coupling Regime}
\label{sec:strong_coupling}

We study in this section the exchange of a quantum excitation in the strong coupling regime, i.e., when the dispersion effect illustrated in Sec.~\ref{sec:physical effects} is not negligible. For simplicity, we consider two specific cases, one for (R1) and another one for (R2).

The idea is to design control fields that remove the dispersion of the excitation onto many energy levels of the oscillators. This can be achieved by refocusing the state onto the desired target state. This process requires necessarily a duration longer than the characteristic time of the transfer between the oscillators. However, the smaller the coupling constant, the longer the physical effect and the better the maximum fidelity. Therefore, the relevant problem is the following: Let $T$ be the time of the transfer, and $g_a'$ be a given value of the coupling constant of the oscillator $a$. In this section, $g_b$ is assumed to be proportional to $g_a$, with a fixed proportionality coefficient. We define: 
\begin{equation}
F_{g_a'}(T) = \max_{g_a\leq g_a'~;~ t\leq T} \left(F[\psi_{target},\psi_0,t,g_a, \vec \Omega(t)= \vec 0]\right).
\end{equation}
Then, is it possible to obtain a control field $\vec \Omega (t)$ such that $F[\psi_{target},\psi_0,T,g_a',\vec \Omega]> F_{g_a'}(T)$ ? If yes, there is an advantage of reaching the strong coupling regime experimentally, but otherwise, it is sufficient to stay in a weak coupling regime. Note that when $g_a'$ is \textit{sufficiently small}, there is a one-to-one mapping between the time where $F[\psi_{target},\psi_0,t,g_a, \vec \Omega(t)= \vec 0]$ is maximum and the value of $g_a'$. For (R1), $T$ is proportional to $g_a^{-1}$, and for (R2), $T$ is proportional to $g_a^{-2}$. For large values of $g_a'$, we have fast and complex Rabi-oscillations, and the maximum fidelity is too small to identify unambiguously a specific time where the excitation is transferred. For this reason, we cannot have a bijection between $T$ and $g_a'$ above a given threshold.

The suitable piecewise constant field shapes for our complex quantum control problems are determined with the GRAPE (GRadient Ascent Pulse Engineering) algorithm~\cite{Khaneja_Optimal_2005}. We refer to Appendix~\ref{sec:numerical_methods} for further details on our application of the GRAPE algorithm. The study is restricted to cases where the transfer works well in the weak coupling regime. Thus, the control fields do not have to reproduce the effective Hamiltonian introduced in Sec.~\ref{sec:effective_hamiltonian}.

\begin{figure}
\includegraphics[width=\columnwidth]{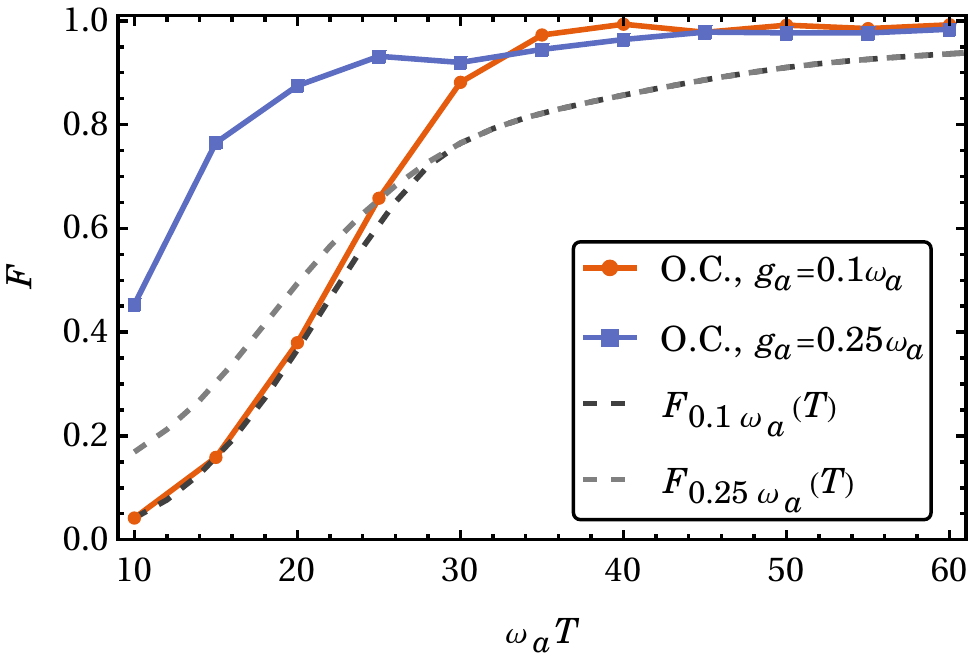}
\caption{Fidelity $F$ with the initial state $\ket{0_S,1_a,0_b}$ and the target state $\ket{-1_S,0_a,1_b}$, for the resonance (R1) and the parameters $\omega_b = 0.3 ~\omega_a, g_b= g_a/\sqrt{2}, \alpha = 0$ (photon-phonon coupling),$\phi_a = \pi/2, \theta_a=0, \gamma_b=0$. The red and blue curves with points correspond to the maximum of fidelity obtained at time $T$ using optimized control (O.C.) fields for two different values of $g_a$. Gray and black dashed curves correspond to $F_{g_a'}(T)$. The parameters of $\hat H_a$, and $\hat H_b$ are the ones obtained in the weak coupling regime (see Sec.~\ref{sec:weak_case1}).} 
\label{fig:fidelity_strong_coupling_1}
\end{figure}

\begin{figure}
\includegraphics[width=\columnwidth]{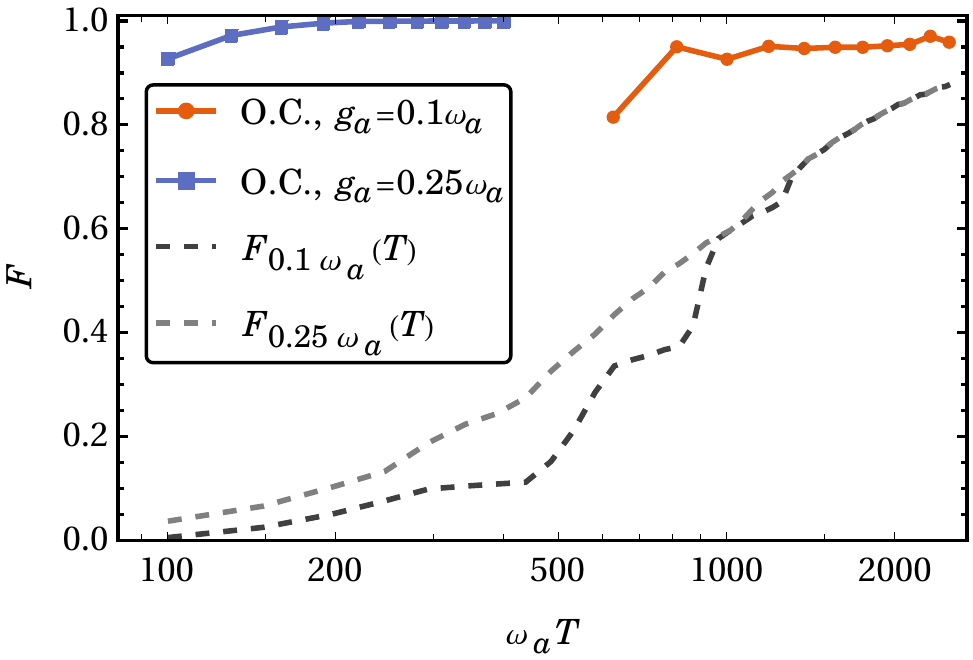}
\caption{
Same as Fig.~\ref{fig:fidelity_strong_coupling_1} but for the resonance (R2) and the parameters $\omega_b = 0.3 ~\omega_a, g_b=g_a/2, \alpha = 1$ (photon-photon coupling),$\phi_a = 0.210573 ~\pi, \theta_a=0, \phi_b = 0.243693~ \pi, \theta_b=0$. The initial state is $\ket{0_S,2_a,0_b}$ and the target state is $\ket{1_S,0_a,2_b}$. The parameters of $\hat H_a$, and $\hat H_b$ are the ones obtained in the weak coupling regime (see Sec.~\ref{sec:weak_case2}).} 
\label{fig:fidelity_strong_coupling_2}
\end{figure}

The problem is explored in Fig~\ref{fig:fidelity_strong_coupling_1} for (R1), with $g_b = g_a/\sqrt{2}$, $\alpha = 0$ (photon-phonon coupling), $\omega_b = 0.3~ \omega_a$, $\theta_a = \gamma_b = 0$, and $\phi_a = \pi/2$. To avoid an alignment of the control with the oscillators $\Omega_x (t)$ is set to $0$. Calculations of optimized controls are made for two values of the coupling parameter $g_a = 0.1~\omega_a$ and $0.25~\omega_a$. 
In the case of $g_a= 0.1~\omega_a$, a fidelity higher than $0.993$ can be achieved in $T=40/\omega_a $, which gives a gain of $13.6\%$. Without control field, a duration $T \approx 185 /\omega_a$ with a coupling $g_a'\approx 0.017 \omega_a$ is necessary to reach a similar fidelity.
Qualitatively similar observations can be made in the case $g_a = 0.25~\omega_a$. Nevertheless, we can stress two main differences: \textit{(i)} we can reach a better fidelity in a shorter time (before $T= 30/\omega_a $), and \textit{(ii)} the maximum fidelity is slightly lower than the one with $g_a = 0.1~ \omega_a$. We interpret this last point with the fact that the convergence of GRAPE can be very slow when it arrives close to the maximum of $F$. Increasing considerably the number of iterations of GRAPE or decreasing the time step of the control discretization can improve slightly the results, but at the price of a considerable increase in the numerical cost.

A similar study is made for (R2), using the following parameters: $\omega_b = 0.3 ~ \omega_a, g_b=g_a/2, \alpha = 1$ (photon-photon coupling),$\phi_a = 0.210573 \pi, \theta_a=0, \phi_b = 0.243693 \pi, \theta_b=0$. Anew, we also set $\Omega_x(t) = 0$. This corresponds to the case already considered in Fig~\ref{fig:R2_WC_2}. Results are plotted in Fig.~\ref{fig:fidelity_strong_coupling_2}. Our observations are the same as the ones made for (R1), but we emphasize that the gain offered by the optimized control field is even more impressive. For example, a fidelity higher than $0.995$ is obtained at $T=190 /\omega_a$, while it is less than $0.1$, at the same time, without control field, and with $g_a'\leq 0.25 ~ \omega_a$. 

\begin{figure*}
\includegraphics[width=\textwidth]{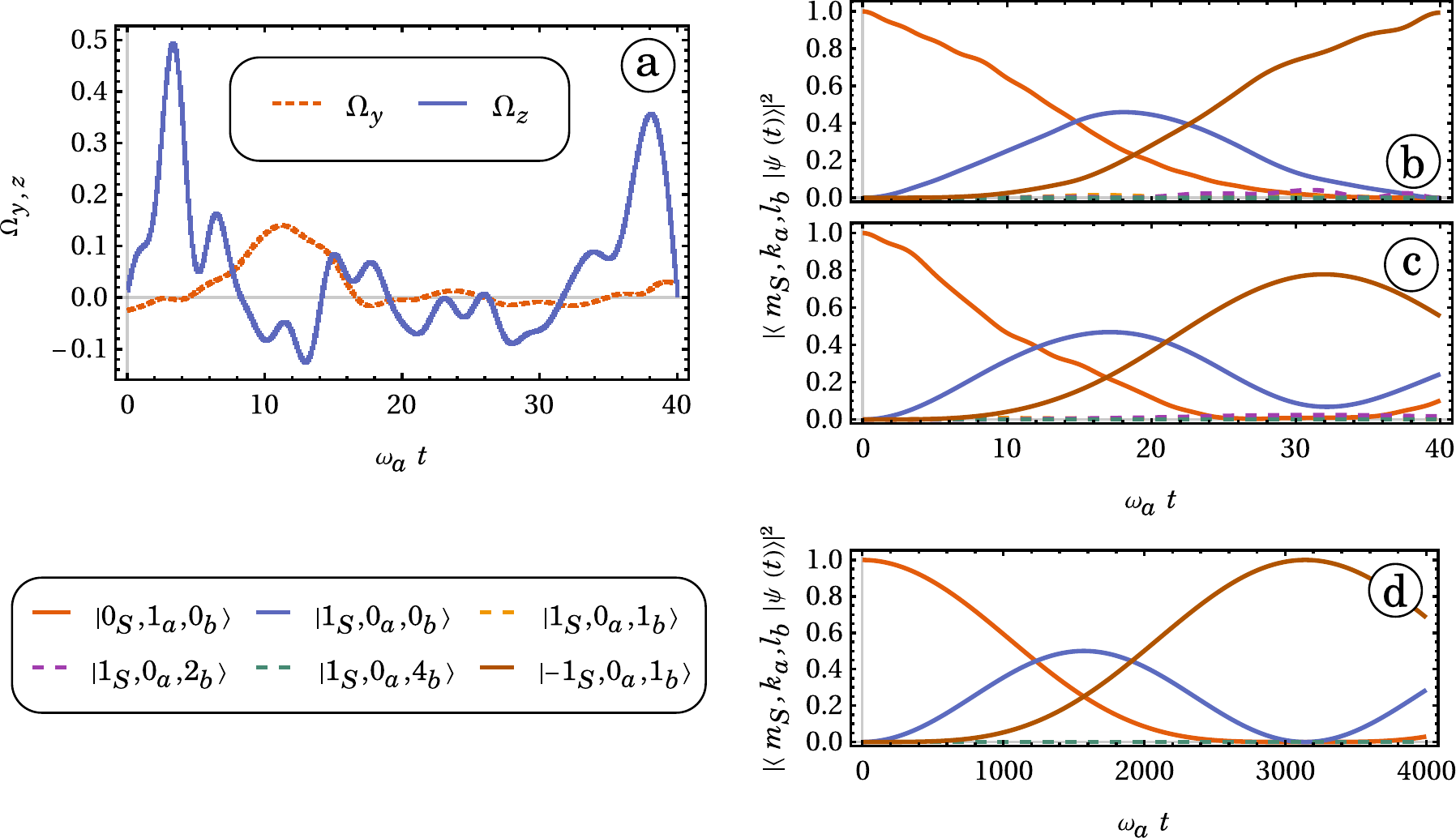}
\caption{\textbf{Panel a)} Optimal control field used to compute the point $F\approx 0.993$ at $T=40/\omega_a$ with $g_a = 0.1 \omega_a$ in Fig.~\ref{fig:fidelity_strong_coupling_1}. \textbf{Panel b)} Time evolution of the population of a few states associated with the control field plotted in panel a). \textbf{Panel c)} Same as panel b), but without a control field. \textbf{Panel d)} Same as panel c), but with the coupling strength $g_a = 0.001 \omega_a$ (weak coupling regime).}
\label{fig:opt_control}
\end{figure*}

In this study, no particular experimental constraints have been imposed on the optimization, to get the ideal limit of the process. The experimental feasibility is analyzed a posteriori. A look at the Fourier transform of the control fields shows that they are made of a few frequencies below $4 ~\omega_a$, with $\omega_a\geq \omega_b$. Additionally, the amplitude of the control field remains rather small. An example of control field is plotted in Fig.~\ref{fig:opt_control} a). With these features, the control fields computed for this study seem realizable as soon as the frequencies of each oscillator are within a domain where we can routinely shape control fields with high accuracy (see e.g., Ref.~\cite{PROBST201942} for an application in electron spin resonance spectroscopy using bismuth donors in silicon). The robustness against variations of the system parameters is another key aspect of the experimental realization. The fields derived for (R1) or (R2) have different robustness. For (R1), a loss of fidelity of $\approx 1\%$ is induced by a variation of the physical parameters $g_a, g_b, D, \omega_z$ about $2$ or $10\%$ (the other parameters are assumed well known up to a sufficiently small precision). For (R2), the same loss of fidelity is induced by variation of the parameters about $0.5$ to $5\%$.

Understanding the physical mechanisms behind the control process is a challenging problem. Our numerical investigations lead us to the conclusion that the process depends both on the resonance and the coupling strength (the shape of the optimal control fields depends strongly on the type of resonance and on the system parameters). The structure of the energy crossings, as illustrated in Fig.~\ref{fig:spectrum} could be responsible for a modification of the control mechanism. However, there is a case easy to analyze: the case (R1) with $g_a =0.1 ~ \omega_a$. The optimized control field of duration $T=40/\omega_a$ is plotted in Fig~\ref{fig:opt_control} a). The trajectory of a few populations associated with this control as well as the population in the case without control are plotted respectively in Fig.~\ref{fig:opt_control} b) and c). An example of trajectory in the weak coupling regime is given in Fig.~\ref{fig:opt_control} d). We remark that the control field forces the state to have a trajectory similar to the one obtained in the weak coupling regime. For this purpose there are two main peaks in $\Omega_z(t)$, located in the intervals $\omega_a t \in[0,10]$ and $[30,40]$ (see the high blue peaks in panel a)). The first one does not have an impressive effect on the dynamics, but it avoids the dispersion of the state in the oscillator $a$. The second part has a similar effect on the oscillator $b$. While this is not visible in the figure, these two peaks are responsible for $\approx 92\%$ of the transfer (this can be observed if we keep only the highest peaks in the control field and we remove the rest). The small fluctuations of $\Omega_z(t)$ on the interval $\omega_a t \in ,30]$, as well as the $\Omega_y(t)$ enhance the control efficiency up to a fidelity of $99\%$. 

\section{Conclusion}
\label{sec:conclusion}

Through this article, we have investigated how two quantum oscillators can exchange one and two quanta of excitation efficiently, using the joint coupling with a three-level system. 
The model system is particularly well adapted to describe quantum phenomena taking place when a solid-state spin interacts with two modes of the electromagnetic field (photons) or one mode of the electromagnetic field, and a vibrational mode modifying the fine structure tensor. Our results are divided into two categories. 

First, we have shown how it is possible to reach a very good fidelity in the transfer of the excitation by engineering the Hamiltonian eigenstates. The exchange of $n$-excitations is maximized when one of the eigenstates is an entangled state with $n$-excitations shared between the two oscillators. We have found that this simple problem has non-trivial solutions, which can be determined analytically or numerically in the weak coupling limit. 

The second category of results concerns the use of a control field on the three-level system to enhance the transfer process when Hamiltonian engineering fails to give us the desired solution.
In particular, a sinusoidal control field can provide enough degrees of freedom to solve the eigenstate problem. Additionally, in the strong coupling regime, a control field can be used to limit the dispersion of the excitation by refocusing the state onto the desired target state. With several examples, we have illustrated the impressive gain of fidelity that can be achieved with this method (e.g., for a 1-excitation exchange, the gain of fidelity is about $14\%$, and for two excitations, the gain can be about $90\%$). Since this improvement takes place in the strong coupling regime (i.e., with short interaction times), controlled quantum excitation transfers are a promising technique to speed up the communication between quantum devices.  

Similar scenarios to the ones explored in this paper have been investigated with adiabatic control methods in the weak coupling regime. For example, in Refs.~\cite{PhysRevA.71.023805}, a STIRAP has been employed for a single quantum exchange between cavities. STIRAP is an adiabatic state-to-state method to transfer an excitation in a three-level system. It requires a very specific form of the perturbed Hamiltonian
, and the process must fulfill adiabatic conditions, such as a long control time. Here, we go beyond this study since the transformations are made at the level of the evolution operator (in the weak coupling regime), we consider perturbed Hamiltonian unsuitable for a STIRAP
, and we work close to the quantum speed limit which is given by the coupling strength (we are not limited to adiabaticity conditions). Moreover, in the strong coupling regime, the STIRAP method cannot work as efficiently as in the weak coupling regime, due to the dispersion effect of the quantum excitation onto many energy levels of the oscillators. This last point has been noted with photons pair production in ultrastrongly coupled matter-radiation systems~\cite{ridolfo_photon_2019}, although the dispersion is not strictly identical to the one of our system.

The proof of concept proposed in this article opens the door to many extensions of this work. In particular, the interaction with the environment may be included, to take into account the experimental constraints of a specific setup. Moreover, the study of one and two-excitation exchanges in the Fock space basis can be extended to other parts of the system spectrum. Other interesting exchanges may be allowed by this system. Finally, a deeper analysis of the evolution operator may give us a general understanding of the control mechanism used to suppress the dispersion of the quantum state.

\begin{acknowledgments}
The authors acknowledge D. L. Shepelyansky for useful discussions. A. D. Chepelianskii has received funding from ANR-20-CE92-0041 (MARS) for this project.
\end{acknowledgments}

\appendix

\section{Diagonalization of the Rabi-Hamiltonian}
\label{sec:rabi}

In this appendix, we revisit several state-of-the-art results concerning the Rabi Hamiltonian in the weak and the strong coupling regime \cite{Xie_2017,Leggett_1987_Dynamics,Chilingaryan_2015}. These quantitative results are helpful to understand qualitatively the physical effects discussed in Sec.~\ref{sec:physical effects}.

While the complexity of the Hamiltonian~\eqref{eq:full_hamiltonian} makes analytic computations arduous and lengthy, the differences between the coupling regimes can be understood with a simplified system composed of one oscillator coupled with the three-level system.
Let us consider that the coupling with the second oscillator is removed, i.e., $g_b = 0$ in Eq.~\eqref{eq:full_hamiltonian}. As a consequence, the basis states of the system Hilbert space are $\ket{m_S,k_a}$, using the same notation as before, but without the oscillator $b$. The following discussion still holds if we set $g_a = 0$ and $g_b \neq 0$. We also left aside the control fields such that $\vec \Omega(t) = \vec 0 ~ \forall t$. Then the Hamiltonian~\eqref{eq:full_hamiltonian} takes the form:
\begin{equation}
\hat H = \hat H_S + \omega_a \Ad \A + g_a \hat H_a (\Ad + \A).
\label{eq:hamiltonian_rabi_1_HO}
\end{equation}
The physics of this system can be studied with the diagonalization of Eq.~\eqref{eq:hamiltonian_rabi_1_HO} in the rotating frame where $\hat H_S$ is canceled.
From the Schr\"odinger equation, we define the evolution operators $\hat U(t) = e^{- \ii t \hat H}$ and $\hat U_S (t) = e^{- \ii t \hat H_S}$, and we determine~\cite{messiah1962quantum}:
\begin{equation}
\begin{split}
\hat U(t) = & \hat U_S (t) \mathbb{T}\exp \left( - \ii \int_0^t dt' \left[ \omega_a \Ad \A \right.\right. \\
& \left. \left. + g_a \hat U_S^{-1}(t')\hat H_a \hat U_S(t') (\Ad + \A)\right] \right),
\label{eq:U_interaction_rabi}
\end{split}
\end{equation}
We can also diagonalize $\hat H_a$, such that $\hat H_a = \hat Q^{-1} \hat S_z \hat Q$. Using this identity, we can rewrite the Hamiltonian in the square brackets of Eq.~\eqref{eq:U_interaction_rabi} into:
\begin{equation}
 \hat U_S^{-1}(t') \hat Q^{-1}\left[ \omega_a \Ad \A + g_a \hat S_z  (\Ad + \A)\right] \hat Q \hat U_S(t').
\end{equation}
The explicit calculation of $\hat Q \hat U_S(t')$ can be lengthy but it is not technically difficult. We focus on the diagonalization of $\omega_a \Ad \A + g_a \hat S_z  (\Ad + \A)$. Following the calculations presented in~\cite{Chen_Numerically_2008}, we deduce the following eigenvalues and eigenstates:
\begin{equation}
E_{m,k} = \omega_a k - \frac{g_a ^2}{\omega_a} m ^2
\label{eq:eigenval_rabi}
\end{equation}
\begin{equation}
\begin{split}
|m_S,k) &= \frac{1}{\sqrt{k!}}\left(\Ad + \frac{g_a}{\omega_a} \hat S_z \right)^k e^{-\frac{g_a^2}{2 \omega_a^2}|\hat S_z|^2
-\frac{g_a}{\omega_a}\hat S_z \Ad} \ket{m_S,0_a} \\
& = e^{-\frac{g_a^2 m^2}{2 \omega_a^2}} \sum_{n=0}^{\infty} \sum_{j=0}^k \frac{(-1)^n \sqrt{k!}}{j! (k-j)! n!} \left( \frac{g_a m }{\omega_a}\right)^{n+k-j} \\
& ~ ~ \times (\Ad)^{j+n} \ket{m_S,0_a}
\end{split}
\label{eq:eigenvec_rabi}
\end{equation}

Note that the eigenstate $|m_S,k)$ depends on the eigenvalue $m$ of $S_z$, hence the notation $m_S$, but the new label $k$ does not depend on the oscillator state alone. It corresponds to the number of excitation created using the operator $\hat A^\dagger = \Ad + \frac{g_a}{\omega_a}\hat S_z$. For this reason, there is no label $a$ under the $k$ of $|m_S,k)$. In the weak coupling regime, we can Taylor expand Eq.~\eqref{eq:eigenvec_rabi} and keep only the highest terms in $g_a/\omega_a$. At order two we have:
\begin{equation}
\begin{split}
&|m_S,k) \approx  \left(1- (2k+1)\frac{g_a^2 m^2}{2\omega_a^2} \right) \ket{m_S,k_a} \\
& ~~  + \frac{g_am}{\omega_a} \left(\sqrt{k} \ket{m_S,(k-1)_a}-\sqrt{k+1}\ket{m_S,(k+1)_a} \right) 
\label{eq:eigenvec_rabi_order_2}
\end{split}
\end{equation}
\begin{figure}
\includegraphics[width=\columnwidth]{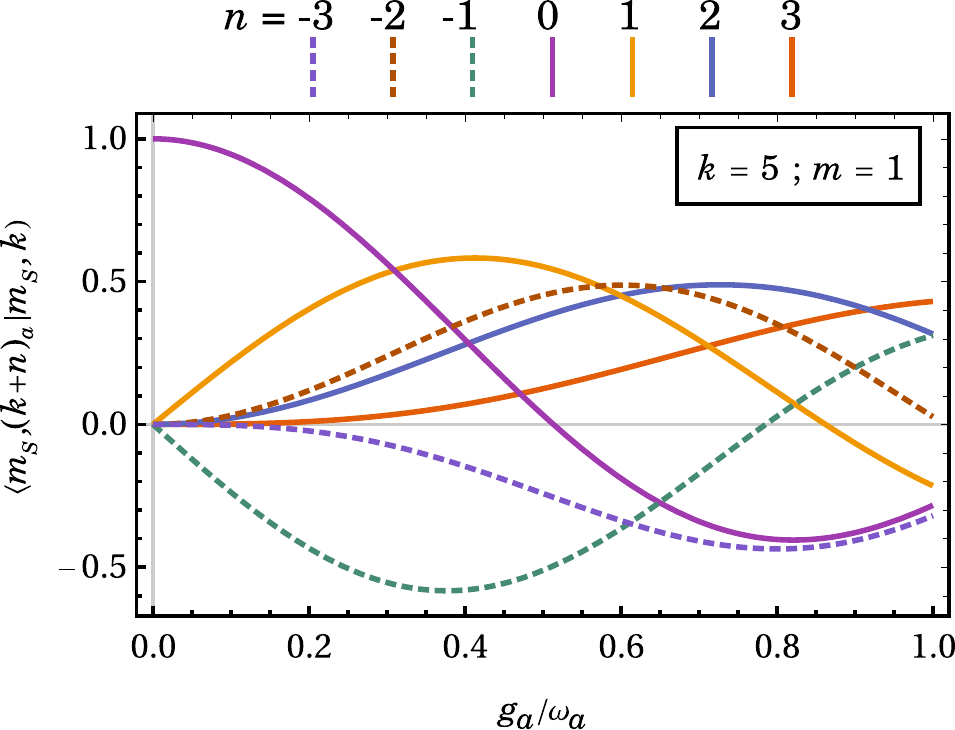}
\caption{Projection of $|m_S,k)$ onto a few basis states $\ket{m_S,(k+n)_a}$ as a function of the parameter $g_a/\omega_a$(see Eq.~\eqref{eq:eigenvec_rabi} for the definition), in the case $k=5$ and $m=1$. For $g_a/\omega_a \lesssim 0.05$ the weak coupling approximation is verified and we recover the approximated state given in \eqref{eq:eigenvec_rabi_order_2}.
}
\label{fig:eigenvalues_rabi}
\end{figure}
With this equation, we see that we can have Rabi oscillations between three states with one or two quantum excitation of difference. This is similar to the rotating wave approximation, which allows us to derive the Jaynes Cummings Hamiltonian~\cite{Jaynes_Comparison_1963} (see~\cite{Irish_Generalized_2007} for a generalization). A qualitative understanding of the strong coupling regime can be derived with an explicit plot of $\bra{m_S,(k+n)_a}m_S,k)$, with $k+n \geq 0$. An example is given in Fig.~\ref{fig:eigenvalues_rabi}. We observe that the larger $g_a/\omega_a$, the higher is the number of state $\ket{m_S,k'_a}$ with a non-negligible projection. As a consequence, if the system is initially in a state $\ket{m_S,k'_a}$, and if the system is in the strong coupling regime, the quantum excitation is going to disperse onto many other states with very different energies of interaction. We will not observe a well-localized exchange of excitation, like in the weak-coupling regime. The situation is similar to the case of a qubit coupled weakly with an ensemble of harmonic oscillators: the excitation is dispersed throughout all the oscillators and it never goes back to the qubit (at least in a reasonable time).

\section{Convergence of the First Order Magnus Expansion}
\label{sec:convergence_magnus}

In this appendix, we analyze the error introduced by the first-order Magnus expansion. To estimate the error, we compute the second-order term of the series, given by:
\begin{equation}
\frac{1}{2} \int_0^t dt_1 \int_0^{t_1}dt_2 [H'(t_1),H'(t_2)],
\label{eq:second_order_magnus}
\end{equation}
where $H'(t)$ is the time-dependent Hamiltonian in the rotating frame (see the middle line of Eq.~\eqref{eq:approx_evol_op_magnus}). We use again the Jacobi-Anger expansion to express $\hat H' (t)$ in a series of the form:
\begin{equation}
\hat H'(t) = \sum_{n=-\infty}^{+\infty} \hat H_n' e^{\ii n \Omega_c t}.
\label{eq:expansion_H'}
\end{equation}
Because $H'$ is a trigonometric polynomial with argument $B= A_c \sin(\Omega_c t)/\Omega_c$, the coefficients $H_n'$ can be decomposed into a sum of terms weighted by Bessel functions $J_n(m A_c/\Omega_c)$, $m \geq 1$ (see Eq.~\eqref{eq:H_a'} and Eq.~\eqref{eq:H_b'} for examples with $m=1,2$). We keep this decomposition implicit for the moment. Inserting Eq.~\eqref{eq:expansion_H'} into Eq.~\eqref{eq:second_order_magnus}, we obtain:
\begin{widetext}
\begin{equation}
\begin{split}
\frac{1}{2} \int_0^t dt_1 \int_0^{t_1}dt_2 [H'(t_1),H'(t_2)] & = \frac{1}{2}\sum_{n=-\infty}^{+\infty}\sum_{k=1}^{+\infty} [H_n',H_{n+k}'] \int_0^t dt_1 \int_0^{t_1}dt_2 ~ e^{\ii ((n+k) t_2 - n t_1)\Omega_c} \\
& =  \frac{1}{2}\sum_{n=-\infty}^{+\infty}\sum_{k=1}^{+\infty} [H_n',H_{n+k}'] \frac{k+n - k e^{-\ii n t \Omega_c} -n e^{\ii k t \Omega_c}}{k n (k+n) \Omega_c^2} .
\end{split}
\label{eq:Jacobi_anger_magnus_order_2}
\end{equation}
We focus on the last fraction in Eq.~\eqref{eq:Jacobi_anger_magnus_order_2}. We remark that it is bounded, except when $n=0$:
\begin{equation}
\left\vert \frac{k+n - k e^{-\ii n t \Omega_c} -n e^{\ii k t \Omega_c}}{k n (k+n) \Omega_c^2} \right\vert \leq \left\lbrace \begin{array}{cr}
\frac{1}{k \Omega_c}\left(\frac{2}{k \Omega_c} +t \right) & ,~n=0 \\ 
\frac{1}{\Omega_c} \left( \frac{1}{k\vert n \vert} + \frac{1}{\vert n(k+n) \vert}  + \frac{1}{k\vert k+n \vert}  \right) & ,~n\neq0
\end{array}  \right. .
\end{equation}
\end{widetext}
In the limit when $\Omega_c \rightarrow \infty$, the leading term is $t/\Omega_c$, and it gives us an estimate of the error made with the first-order Magnus expansion.
The larger $\Omega_c$, the better the approximation is. Nevertheless, this error estimate can be very large, compared to the true error. To underline this point, 
we consider only the leading term in Eq.~\eqref{eq:Jacobi_anger_magnus_order_2}, and we further assume that: $[H_0',H_1'] \propto J_0\left(m_1 \frac{A_c}{\Omega_c}\right) J_1\left(m_2 \frac{A_c}{\Omega_c}\right)$, $m_1,m_2 \geq 0$. Then, we have:
\begin{equation}
\begin{split}
&\left\Vert \frac{1}{2} \int_0^t dt_1 \int_0^{t_1}dt_2 [H'(t_1),H'(t_2)] \right\Vert  \\
&~ \sim ~ C \frac{t}{\Omega_c} \left\vert J_0\left(m_1 \frac{A_c}{\Omega_c}\right) J_1\left(m_2 \frac{A_c}{\Omega_c}\right)\right\vert,
\end{split}
\end{equation}
with $C$ a constant that does not depend on $m_1$, $m_2$,  $t$, $A_c$, and $\Omega_c$. The absolute value of the product of $J_0 J_1$ is in general very small. For example, the highest maximum of $\vert J_0(x)J_1(x)\vert$ is $\approx 0.339$ for $x \approx 1.082$, the second one is $\approx 0.076$ at $x \approx 4.620$. We can thus expect that in many situations, the error is reduced by the small values taken by the Bessel functions.

The fact that the approximation works very well for a rather small value of $\Omega_c$ (see Sec.~\ref{sec:effective_hamiltonian}) is, however, not guaranteed and it is verified a posteriori. The same approximation applied to another system may require much larger values of $\Omega_c$ (for example, see~\cite{janowicz_non-markovian_2000}).

\section{Notes on the numerical computations}
\label{sec:numerical_methods}

In this section, we provide several key elements concerning numerical simulations and optimizations.   

Simulations are performed using two truncated Fock spaces. The number of energy levels is chosen sufficiently large to avoid any population transfer in the highest excited states. This is a necessary condition to limit numerical artifacts. In the simulations presented in the main text, we have kept between 6 and 16 energy levels. The respective dimensions for the total Hilbert spaces are then $6^2*3 = 108$ and $16^2*3=768$.  All control fields are defined with piecewise constant functions. The time step $\Delta t$ is adapted to the minimum resolution required for the control field. The numerical integration of dynamics is performed with matrix exponential: $\ket{\psi (n \Delta t)} = \mathbb{T} \prod_{k=0}^{n-1} e^{-\ii \Delta t~ H(k\Delta t)}\ket{\psi_0}$. Note that the Hamiltonian operator is constructed with tensor products of truncated creation and annihilation operators expressed in the $\ket{m,k,l}$ basis. This way of constructing the operator is numerically very fast and it is advantageous when we need to optimize the parameters of the Hamiltonian. However, this method can lead to numerical artifacts if the coupling constants $g_a$ and $g_b$ are too high, or if the truncated space is too small. Errors can be reduced using the diagonal basis of the Rabi-Hamiltonian~\cite{Chen_Numerically_2008} defined in Eq.~\eqref{eq:eigenval_rabi} and \eqref{eq:eigenvec_rabi}, but at the price of a longer computation time. However, in all our cases study, the couplings are sufficiently small and the states remain sufficiently close to the ground state to avoid any significant numerical artifacts.  

Two types of numerical optimizations are performed in this study. The optimization of the parameters of the Hamiltonian, or the optimization of a control field.

In the first case, optimizations are made using the built-in function of Mathematica: NMinimize \cite{NMinimize}. This function allows us to perform global optimizations, and the convergence is generally excellent with a few variables on a bounded domain. This function is used with the option ``SimulatedAnnealing''.

The optimizations of control fields are performed using the GRAPE algorithm \cite{Khaneja_Optimal_2005,Boscain_Introduction_2021}. This algorithm is an efficient gradient-based optimization algorithm to solve quantum control problems. Optimizations with GRAPE require the input of an initial control field which is iteratively deformed to converge towards a local optimum of the objective function (here, the fidelity $F$). The initial control field is taken to be a sinusoidal function of very small amplitude such that $\vec \Omega(t) \approx \vec 0$. This choice is motivated by the fact that without control, we can reach approximately the target state, and we look for a solution that allows us to improve the free evolution without introducing too much energy into the system. As illustrated by our results (see the main text), excellent convergences are observed, and we can hope to be close to the optimal solution. The fields obtained after optimization do not depend significantly on the initializing field. However, it is not excluded that other solutions may be found with initial fields whose energy is noticeably larger than zero.  

\section{Controllability}
\label{sec:controllability}

In this appendix, we address a few comments on the system's controllability. Controllability checks, when they are doable, are generally necessary to ensure the existence of a solution to the control problem.

A quantum system is said \emph{pure state controllable}~\cite{Albertini_Notions_2001} if for every pair of initial and target states $\ket{\psi_0}$, $\ket{\psi_{target}}$, there exists a piecewise constant control field $\Omega$ such that $\ket{\psi_{target}}=\hat U(T,\Omega) \ket{\psi_0}$, $T>0$. This is a very strong result which is usually not achieved with infinite dimensional systems. In this case, we use the notion of approximate controllability. A quantum system is said \emph{approximately controllable}~\cite{illner2006limitations} if for every pair of initial and target states $\ket{\psi_0}$, $\ket{\psi_{target}}$, for every $\epsilon>0$ there exists $T_\epsilon>0$ and a piecewise constant control field $\Omega$ such that $\Vert \ket{\psi_{target}}-\hat U(T_\epsilon,\Omega) \ket{\psi_0}\Vert < \epsilon$. This means that we can reach any state with an arbitrarily small error $\epsilon$. The time of control is (in general) dependent on the error. Note that Markovian interactions with an environment lead to a loss of controllability.

Proving the controllability of a quantum system is in general a difficult task, specifically for an infinite quantum system like the one considered in this paper. Such a result is out of the scope of this paper. However, the control tasks we are interested in are non-trivial, and a minimum of knowledge on this issue is necessary. We can verify that the system with truncated Fock spaces is controllable. By increasing the dimension of the truncated Hilbert space, we can verify if the controllability is kept. This does not give us a mathematical proof, but at least, it gives us a first clue in this direction. The controllability of a finite-dimensional system can be determined by computing the dimension of the dynamical Lie algebra. If it is equal to $\text{dim} \Mc H ^2-1 = \text{dim}\left[ \mathfrak{su}(\text{dim} \Mc H) \right]$, then the system is controllable. This can be achieved numerically with the algorithm presented in Ref.~\cite{schirmer_complete_2001}. Since the calculation time grows considerably with the size of the Hilbert space, we are restricted to truncated systems with the smallest dimensions. We have verified that the Hamiltonian \eqref{eq:full_hamiltonian} is controllable in the case $\theta_a = 0,\phi_a=\pi/2, \alpha = 0, \gamma_b = 0$ up to 3 energy levels of the oscillators. Moreover, we have verified that the Rabi Hamiltonian \eqref{eq:hamiltonian_rabi_1_HO} with $\hat H_a =\hat S_x$ and controls on $\hat S_y$ and $\hat S_z$ is controllable up to 5 energy levels of the oscillator. In addition to these results, we can emphasize that the Rabi-Hamiltonian is approximately controllable with a coherent control of the bosonic mode~\cite{Boscain_control_2015}. Based on these results, we conjecture that our model system \eqref{eq:full_hamiltonian} is controllable.

\bibliographystyle{apsrev4-2}

\end{document}